\documentclass{article}

\usepackage[eandd,final]{neurips_2026}
\usepackage[utf8]{inputenc}
\usepackage[T1]{fontenc}
\usepackage{hyperref}
\usepackage{url}
\usepackage{booktabs}
\usepackage{amsfonts}
\usepackage{amsmath}
\usepackage{nicefrac}
\usepackage{microtype}
\usepackage{graphicx}
\usepackage{multirow}
\usepackage{enumitem}
\usepackage{capt-of}
\usepackage[table]{xcolor}
\usepackage{xspace}
\usepackage{placeins}

\title{Who Called? V33DA: A Physically Verified Multimodal Benchmark for Vocal Attribution in Zebra Finch Groups}

\author{
  Maris Basha$^{1}$\quad
  Yuhang Wang$^{1}$\quad
  Xiaoran Chen$^{2}$\quad
  Longbiao Cheng$^{1}$\quad \\
  Luca Yapura$^{1}$\quad
  Anja T. Zai$^{1}$\quad
  Mathieu Salzmann$^{2}$\quad
  Richard Hahnloser$^{1}$ \\[4pt]
  $^{1}$Institute of Neuroinformatics, University of Zurich and ETH Zurich \\
  $^{2}$Swiss Data Science Center and CVLab, EPFL \\[6pt]
  \small
  \mbox{\href{https://huggingface.co/datasets/songbirdini/V33DA}{\raisebox{-1pt}{\includegraphics[height=10pt]{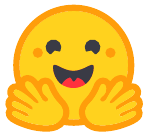}}~Dataset}}
  \quad
  \mbox{\href{https://github.com/marisbasha/v33da}{\raisebox{-1pt}{\includegraphics[height=10pt]{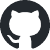}}~Code}}
}

\begin{document}

\maketitle

\vspace{-1.5em}
\begin{center}
\includegraphics[width=0.95\textwidth]{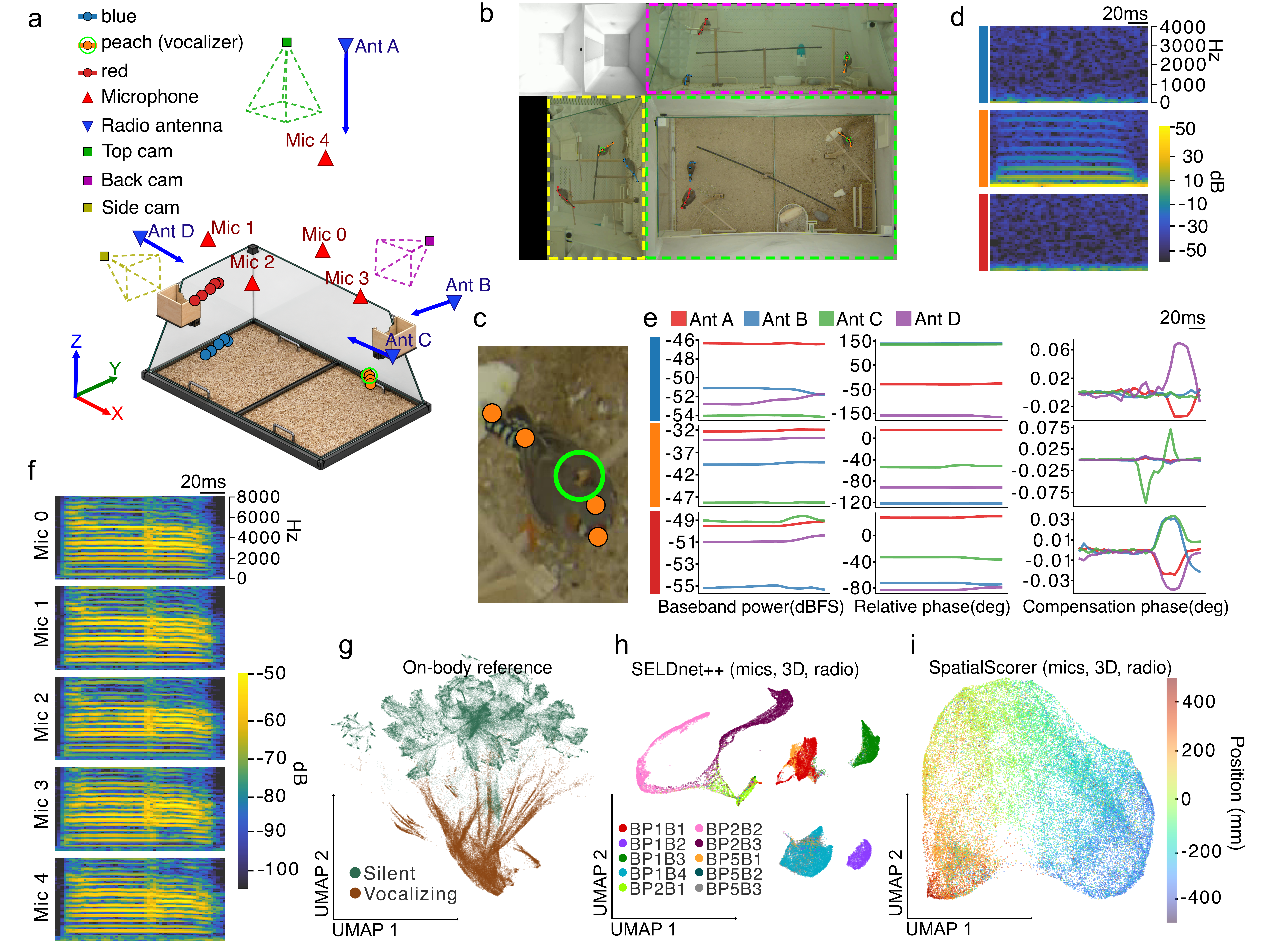}
\captionof{figure}{\textbf{V33DA overview and representation geometry.} \textbf{a.}~BirdPark rendering with calibrated microphones, camera poses, the 4-antenna FM receiver array, and triangulated bird positions. \textbf{b.}~Synchronized multi-view frame with tracked individuals. \textbf{c.}~Candidate-centered crop; the neon-green ring marks the vocalizing bird's backpack. \textbf{d.}~Accelerometer-derived vibration spectrograms; left color bars mark the blue, peach, and red birds. \textbf{e.}~Per-antenna (A--D) radio telemetry---baseband signal power, relative phase, and compensation phase---over the event, with the same color bars. \textbf{f.}~Five-channel microphone spectrograms. Panels \textbf{a--f} show the same verified event across geometry, video, on-body sensing, radio telemetry, and multichannel audio. \textbf{g.}~UMAP of withheld on-body reference embeddings: the reference cleanly separates the caller from silent birds. \textbf{h.}~SELDnet++ embeddings organized primarily by bird identity. \textbf{i.}~SpatialScorer embeddings organized primarily by vocalizer position. Panels \textbf{g--i} preview the main result: transferable attribution depends on spatial organization more than on identity recognition.}
\label{fig:overview}
\end{center}

\begin{abstract}
Deciding which member of an animal group produced a vocalization and where that vocalization was produced are central to studying social communication but are rarely evaluated with physically verified ground truth. Existing resources either localize sounds without identifying the caller, or recognize individuals from isolated vocalizations without reliance on the 3D candidate geometry. We close this gap with \textsc{V33DA}, a dataset-centered multimodal benchmark for spatial vocal attribution in social zebra finches: it comprises 33{,}625 verified events coupling 5-channel audio with three camera views, calibrated 3D pose for every visible candidate, and per-bird radio telemetry across 10 individuals and three experiments. Caller identity is verified from an on-body accelerometer-derived vibration signal; this channel is withheld from benchmark models and used only by an on-body reference ceiling. We also provide \textsc{V33DA++}, an unevaluated companion release with verified overlapping-call events and $\pm 2$~s context windows, intended to support future source-separation and vocal-activity-detection studies.

V33DA separates familiar-individual recognition from transferable spatial attribution. Given a detected vocalization and the visible candidate birds, the task is to select the caller, with 3D source localization as a secondary diagnostic. We evaluate it under three regimes: session-disjoint testing on familiar birds from held-out recording dates, held-out-experiment transfer to a new cohort, and leave-one-individual-out evaluation on a bird whose vocalizations are held out from training. A broad set of reference methods exposes a consistent gap between in-domain accuracy and transfer to unseen individuals: methods that can exploit familiar vocal identity perform well on known callers but degrade sharply on unseen ones, whereas candidate-aware methods preserving explicit spatial reasoning transfer substantially better. The on-body reference stays near-perfect across regimes, confirming the internal consistency of the accelerometer-verified labels and quantifying the ceiling available when the withheld verification channel is observed. We release V33DA with fixed attribution/localization protocols and V33DA++ with overlap and long-context files, together with code for the evaluated methods.

\end{abstract}


\section{Introduction}
\label{sec:introduction}
Vocal communication in a social group is only interpretable at the level of individuals once each call can be assigned to the individual that produced it. In zebra finches, vocalizations carry information about identity, pair bonds, and social context~\citep{elie2016vocal,elie2018zebra}; without caller attribution, these signals cannot be linked back to the animals that generated them. The problem is therefore the entry point to studying who communicates with whom, when, and under what social conditions.

Caller attribution in freely behaving groups is difficult because the evidence is distributed across modalities. The sound arrives at several microphones, the candidate animals move through a shared 3D space, and visual identity, posture, proximity, and acoustic structure all change over time. A model can also solve the wrong problem: if the same individuals appear in training and test, high accuracy may reflect recognition of familiar voices rather than a transferable ability to assign a call to the correct visible animal. A useful benchmark must therefore separate familiar-individual recognition from spatial attribution that generalizes to new individuals.

Existing resources capture important pieces of this problem but not the full setting. Large bioacoustic corpora support species, call-type, or individual-recognition tasks, but typically do not provide the synchronized candidate geometry needed to decide which animal in a group produced a sound~\citep{kahl2021birdnet,hagiwara2023beans,rauch2025birdset}. Spatial-audio benchmarks evaluate localization or direction-of-arrival estimation, but usually do not provide physically verified caller identity for freely behaving animals~\citep{evers2020locata,politis2022dcase,shimada2023starss}. Active-speaker and diarization datasets ask related attribution questions in humans, but rely on perceptual annotation, face or mouth cues, or close-talk references, and rarely provide calibrated 3D candidate state~\citep{roth2020ava,xu2022avaavd,nguyen2025unitalk}. On-body animal-recording studies can verify who called, but they do not by themselves define a public multimodal benchmark for testing transfer across unseen individuals. What is missing is a dataset that combines physical verification, synchronized group state, and evaluation splits designed to expose identity shortcuts.

We introduce \textsc{V33DA} (\emph{Vocal 3-Experiment 3D Attribution}), a dataset-centered multimodal benchmark for spatial vocal attribution in social zebra finches, built on the BirdPark recording system~\citep{ruttimann2024birdpark}. Each of the 33{,}625 released events includes 5-channel audio, three calibrated camera views, per-frame 3D pose for every visible candidate, per-bird radio telemetry, and an on-body accelerometer-derived vibration channel used for verification. The verification channel is withheld from benchmark models and used only to define an on-body reference ceiling. We additionally provide V33DA++ as an unevaluated companion release: 3,784 verified overlapping-call events excluded from the main V33DA benchmark and $\pm 2$~s padded context windows. These files are intended to support future work on source separation, vocal activity detection, and longer-context active-speaker tasks; no result in this paper depends on them.

The primary benchmark task is candidate-conditioned caller attribution: given a detected vocalization and the currently visible candidate birds, choose which bird produced the call. We also report 3D source localization as a diagnostic of whether a method has learned usable spatial evidence rather than only a candidate ranking. Evaluation uses three regimes: session-disjoint testing on familiar birds from held-out recording dates, leave-one-bird-out testing on a bird whose vocalizations are held out under the same cage and sensor setup, and held-out-experiment testing on a new cohort with no birds shared between train and test. Together, these regimes distinguish in-domain reuse, unseen-vocalizer generalization under fixed acquisition conditions, and broader experiment-level transfer.

The paper makes three contributions:
\begin{itemize}[nosep,leftmargin=*]
    \item \textbf{A physically verified multimodal release.} V33DA provides synchronized audio, video, 3D pose, radio telemetry, and withheld on-body verification for social vocal attribution.
    \item \textbf{Splits that separate familiar from unseen individuals.} The evaluation protocol separates familiar-individual testing from unseen-individual and held-out-experiment transfer, with fixed splits and common attribution/localization metrics.
    \item \textbf{A benchmark lesson about shortcut learning.} Reference methods show that high in-domain accuracy can come from familiar-identity cues, while transfer to unseen individuals depends much more strongly on candidate-aware spatial reasoning.
\end{itemize}

We release the dataset, evaluation code, and implementations of the evaluated methods.


\section{Related Work}
\label{sec:related}

\paragraph{Task landscape.}
Several tasks touch the problem of assigning vocalizations to individuals, but they differ in what is given and what must be predicted. \emph{Individual vocal identification} (ID)~\citep{elie2018zebra,zuercher2023caller,nagrani2020voxceleb} starts from an isolated vocalization and predicts the identity of a known caller; the scene and the other possible callers are usually absent. \emph{Sound-source localization} (SSL)~\citep{knapp1976gccphat,evers2020locata,hyvl2023} starts from array audio and predicts a position or direction of arrival; it can recover where the sound came from without deciding which animal at that position called. \emph{Source separation} (Sep) and \emph{vocal activity detection} (VAD)~\citep{ryant2019dihard,stoeger2024meerkat} operate over longer audio: they ask which sources are active, when they are active, and sometimes how to separate their waveforms, but they do not necessarily attach each event to a tracked individual. \emph{Active-speaker detection} (ASD) and \emph{audio-visual diarization} (AVD)~\citep{roth2020ava,xu2022avaavd,chung2020voxconverse} ask which visible person is speaking over time, most often in human video where faces, mouths, or close-talk microphones provide supervision.

V33DA targets \emph{candidate-conditioned caller attribution} (CA). The event has already been detected, the candidate animals visible at that moment are known, and the model must choose which candidate produced the call. This is different from ID because the test caller may be unfamiliar, different from SSL because the output must be an individual rather than only a point in space, and different from tracking because the candidates are already tracked. The task therefore needs both verified caller identity and synchronized scene geometry: recognizing a familiar voice is not enough under unseen-individual transfer, and localizing a sound is not enough unless the location can be linked to a candidate animal. Appendix Table~\ref{tab:task_coverage} maps these tasks onto existing resources.

\paragraph{Dataset families.}
Existing resources cover different parts of this task landscape. Bioacoustic recognition datasets, models, and benchmarks provide species, call-type, or individual labels at scale, but generally lack synchronized group geometry: they can tell whether a vocal signature identifies an individual, not which visible animal in a shared scene produced a particular event~\citep{kahl2021birdnet,hagiwara2023beans,rauch2025birdset,elie2018zebra,zuercher2023caller,birdaves2025,basha2024vocsim}. Spatial-audio and sound event localization and detection (SELD) datasets provide localization or direction-of-arrival targets, including real and simulated array recordings, but usually do not provide physically verified animal caller identity~\citep{politis2022dcase,evers2020locata,guizzo2022l3das,shimada2023starss}. Human active-speaker and diarization datasets ask related attribution questions, but rely on human scenes, perceptual labels, face or mouth cues, or close-talk references rather than calibrated 3D state for all candidate animals~\citep{roth2020ava,xu2022avaavd,nguyen2025unitalk,carletta2005ami,grauman2022ego4d,ryant2019dihard,chung2020voxconverse}. On-body animal-recording studies physically verify callers, but they are typically not public multimodal benchmarks combining room audio, synchronized video, 3D pose, radio telemetry, and transfer splits~\citep{anisimov2014reconstruction,gill2016backpack,stoeger2024meerkat,music2022rookognise,prat2017annotated}.

\begin{minipage}{\textwidth}
\centering
\scriptsize
\setlength{\tabcolsep}{2.5pt}
\captionof{table}{\textbf{Comparison of V33DA with related datasets.} Task abbreviations are defined in the text (classif.\ = classification, diar.\ = diarization, verif.\ = speaker verification). \textbf{N}: number of individuals; \textbf{Conc.}: individuals recorded concurrently; \textbf{Mics}: microphones; \textbf{Vis.}: synchronized video ($\circ$: in-the-wild video without calibrated geometry); \textbf{Verif.}: source of the caller label (on-body accel.\ = accelerometer on the animal; body-mounted mic = microphone on the animal; perceptual = human annotation); vocs = vocalizations, utt.\ = utterances, segs = segments.}
\label{tab:comparison}
\resizebox{\textwidth}{!}{%
\begin{tabular}{llllrrrcll}
\toprule
 & \textbf{Domain} & \textbf{Task} & \textbf{Size} & \textbf{N} & \textbf{Conc.} & \textbf{Mics} & \textbf{Vis.} & \textbf{Verif.} \\
\midrule
\rowcolor{black!12}\multicolumn{9}{@{}l}{\textit{\textbf{Physically verified animal caller attribution and localization}}} \\
Anisimov+ \citeyearpar{anisimov2014reconstruction} & Zebra finch & CA          & methods demo            & 9     & 9       & 1/bird   & --        & on-body accel. \\
Gill+ \citeyearpar{gill2016backpack}               & Zebra finch & CA          & 182k calls              & 8     & 8       & 1/bird   & --        & body-mounted mic \\
Rookognise \citeyearpar{music2022rookognise}       & Rook        & CA + ID     & 17.7k vocs              & 15    & 15      & 6        & --        & obs. + MC audio \\
MeerKAT \citeyearpar{stoeger2024meerkat}           & Meerkat     & SSL$_{\text{pre}}$ + classif. & 251k events & 41 & --   & 1/collar & --        & body-mounted mic \\
Prat+ \citeyearpar{prat2017annotated}              & Fruit bat   & repertoire  & 90k events              & 33    & 2--8    & 1/chamb. & \checkmark & mouth on IR video \\
HyVL \citeyearpar{hyvl2023}                        & Mouse       & SSL         & 13.7k USVs              & 18    & 1--3    & 68       & \checkmark & beamforming + pose \\
VCL \citeyearpar{vcl2024}                          & Rodent      & SSL         & 767k events             & --    & 1--2    & 4--24    & \checkmark & video pose / earbud \\
\midrule
\rowcolor{black!12}\multicolumn{9}{@{}l}{\textit{\textbf{Individual vocal recognition}}} \\
Elie+ \citeyearpar{elie2018zebra}                  & Zebra finch & ID          & 8k+ calls               & 45    & 1       & 1        & --        & isolated rec. \\
Phaniraj+ \citeyearpar{zuercher2023caller}         & Marmoset    & ID          & 7.1k calls              & 20    & 1       & 1        & --        & isolated rec. \\
VoxCeleb \citeyearpar{nagrani2020voxceleb}         & Human       & ID + verif. & 1M+ utt.                & 7363  & 1       & 1        & $\circ$   & SyncNet + face recog. \\
\midrule
\rowcolor{black!12}\multicolumn{9}{@{}l}{\textit{\textbf{Human active-speaker, diarization, and spatial audio}}} \\
AVA-ActSp. \citeyearpar{roth2020ava}               & Human       & ASD         & 29.6k turns             & --    & 1--3    & 1        & $\circ$   & perceptual \\
AVA-AVD \citeyearpar{xu2022avaavd}                 & Human       & AVD         & 40.3k segs              & 1500  & 2--24   & 1        & $\circ$   & perceptual \\
UniTalk \citeyearpar{nguyen2025unitalk}            & Human       & ASD         & 44.5~h                  & 48693 & 2--3    & 1        & $\circ$   & perceptual \\
AMI \citeyearpar{carletta2005ami}                  & Human       & ASR + diar. & 93.6k segs              & 189   & 4       & 8 + close-talk & \checkmark & close-talk \\
DIHARD II \citeyearpar{ryant2019dihard}            & Human       & diar.       & 386 recs                & --    & 3--7    & 1        & --        & perceptual \\
VoxConverse \citeyearpar{chung2020voxconverse}     & Human       & AVD         & 27.7k turns             & --    & 2+      & 1        & $\circ$   & SyncNet + perceptual \\
STARSS23 \citeyearpar{shimada2023starss}           & General     & SELD        & 17.6k events            & 57    & varies  & 4 (FOA/MIC) & \checkmark & mocap \\
\midrule
\rowcolor{black!5}
\textbf{V33DA (ours)} & \textbf{Zebra finch} & \textbf{CA + SSL + ID} & \textbf{33.6k events} & \textbf{10} & \textbf{3--4} & \textbf{5} & \textbf{\checkmark} & \textbf{on-body accel.} \\
\bottomrule
\end{tabular}%
}
\end{minipage}

Multi-animal vision datasets provide the candidate state that attribution needs but not the vocal evidence itself. DeepLabCut, SLEAP, and idTracker anchor 2D pose and identity tracking~\citep{mathis2018deeplabcut,pereira2022sleap,romeroferrero2019idtracker}; 3D-POP and 3D-MuPPET extend to multi-view 3D skeletons~\citep{naik2023threedpop,naik2024threedmuppet}; and PAIR-R24M, MammAlps, and CalMS21 scale up pairwise or social-behavior annotation~\citep{dunn2021pairr24m,mammalps2025,sun2021calms21}. These resources are essential for reconstructing who is where, but caller attribution asks a different question: which of those candidates produced the audio event?

\paragraph{Radio and transfer.}
Multi-antenna radio for spatial reasoning is well-established outside bioacoustics: cellular and 5G exploit angle-of-arrival, time-difference-of-arrival, and channel-state information for meter- to centimeter-scale positioning~\citep{delperalrosado2018survey,wymeersch2017mmwave}, and WiFi/mmWave channel-state sensing carries the same principles to indoor localization, gesture recognition, and presence detection~\citep{ma2019wifi}. To our knowledge no prior public animal-vocalization dataset ships per-individual radio telemetry alongside calibrated position and audio-visual channels. Separating in-domain generalization from distribution shift is standard in transfer benchmarks such as DomainBed and WILDS~\citep{gulrajani2021domainbed,koh2021wilds}; V33DA brings the same logic to social vocal attribution by separating session, leave-one-bird-out, and held-out-experiment regimes.

\paragraph{V33DA's position.}
What remains missing is a public benchmark that brings these ingredients into the same event: synchronized state for every visible candidate, physically verified caller labels, and evaluation splits that test whether attribution transfers beyond familiar individuals. V33DA combines 5-channel audio, multi-view video, per-candidate 3D pose, per-bird radio telemetry, and an on-body verification channel that is withheld from benchmark models. Related resources support parts of this protocol but not the same verified evaluation: STARSS23 has no tracked animal candidates or caller identities, Rookognise has no synchronized trajectories or calibrated 3D geometry, and VCL has no independently verified dyadic caller labels. HyVL comes closest, but it assigns callers by matching acoustic localization to tracked snouts, so localization-first methods evaluated on it would be scored against localization-derived labels; the withheld on-body channel avoids this circularity. With 33{,}625 verified vocalizations, V33DA is also larger than Rookognise (about 17.7k) and HyVL (about 13.7k). Table~\ref{tab:comparison} and Appendix Table~\ref{tab:task_coverage} summarize how this combination differs from prior resources.

\section{Dataset}
\label{sec:dataset}

\paragraph{Recording setting.}
All recordings come from the BirdPark aviary~\citep{ruttimann2024birdpark}, a synchronized multimodal installation for observing small social groups under naturalistic interaction. The benchmark spans three experiments sharing the same sensing configuration, each housing 3--4 zebra finches. Five calibrated microphones at known 3D positions record at 24{,}414~Hz, three synchronized cameras (top, back, side) capture video at 47.68~fps, and 3D pose is triangulated from 5 keypoints per bird. A 4-antenna receiver array provides per-bird radio telemetry, and an FM-demodulated on-body signal exposes thoracic vibration during vocalization~\citep{eisenring2022accelerometer}, the channel that anchors our caller labels. Figure~\ref{fig:overview}(a--f) shows one released event across these aligned views.

\paragraph{Label construction and manual validation.}
Caller labels are anchored in the on-body vibration signal rather than in perceptual judgment. For each bird, a multichannel WhisperSeg model~\citep{gu2024whisperseg} proposes candidate windows from the demodulated on-body trace, and a sample is retained only when the event is visible in the microphone channels, the labeled bird shows the corresponding on-body vibration signature, and no overlapping call from another bird is present. Candidate state (detections, backpack-color identity, and triangulated 3D pose) is reconstructed from the synchronized video and radio streams (Appendix~\ref{app:detection}, \ref{app:identity}, \ref{app:triangulation}), and every candidate vocalization is manually reviewed against the full multimodal evidence (Appendix~\ref{app:human}). The on-body vibration channel physically determines which bird produced each call; manual review therefore checks that the room audio, video, and 3D pose are consistent with that label, rather than inferring caller identity perceptually as in human active-speaker datasets. Of 39{,}329 reviewed candidates, 1{,}768 are rejected as invalid; reprojection filtering removes a further 152, and overlap filtering removes 3{,}784, yielding 33{,}625 released single-vocalizer events.

\paragraph{Release scope.}
The 33{,}625 events span BP01, BP02, and BP05, with 16{,}219 / 11{,}577 / 5{,}829 events per experiment (Appendix Table~\ref{tab:dataset}). All 10 released birds are distinct physical individuals, and no bird appears in more than one experiment. Birds are referenced by per-experiment backpack color codes, so a color reused across experiments refers to different physical individuals. Each event carries a per-event \texttt{vocalizer\_idx} $\in \{0,\dots,3\}$ identifying the caller within the event's up-to-four visible candidates; this index is stable within an experiment but experiment-local (full schema: Appendix~\ref{app:pipeline}). Audio clips range from 40 to 784~ms (mean 150~ms, median 125~ms), and each sample includes 4--38 video frames with aligned radio telemetry.

\paragraph{V33DA++.}
V33DA++ is an unevaluated companion release to the main V33DA benchmark and ships three buckets; no result in this paper depends on it. \texttt{overlap} contains reviewer-verified calls dropped from V33DA because another bird vocalized in the same window, each tagged with the concurrent callers verified from their on-body vibration channels. \texttt{padded} contains $\pm$2~s windows around V33DA events with original onsets and offsets preserved, supplying long-range context without altering the underlying event. \texttt{overlap\_padded} applies the same $\pm$2~s context to overlap events for long-context multi-caller tasks.


\section{Benchmark Design}
\label{sec:benchmark}

The main task is candidate-conditioned caller attribution. We evaluate every model under three progressively harder splits, designed so a model that has only memorised the training birds cannot look strong on all of them.

\paragraph{Evaluation regimes.}
\label{sec:eval_regimes}
\textbf{Session-disjoint} separates train, validation, and test by recording date while keeping all birds in every partition. This is the in-domain setting: a model sees every bird at training time and is asked to generalise across recording sessions of the same individuals, so scoring well requires nothing more than recognising familiar voices. \textbf{Held-out experiment} rotates the test experiment across BP01, BP02, and BP05 and trains on the remaining two. Because no bird appears in more than one experiment, every test event comes from an individual the model has never seen, and the shift also carries realistic cohort and recording-date differences. \textbf{Leave-one-bird-out} holds out positive vocalization examples for one bird within a single experiment while the remaining birds provide training and validation, and the cage, microphone geometry, and recording dates stay fixed. The held-out bird may still appear as a silent visible candidate in other birds' events, where it is always labeled as a non-vocalizer, so LOO tests held-out-vocalizer generalization under fixed sensing conditions together with resistance to this negative prior; held-out experiment, where no bird is shared, is the cleaner transfer regime. This yields 4 folds for BP01 and 3 folds each for BP02 and BP05, for 10 folds in total. Table~\ref{tab:regimes} summarises what varies, what is held fixed, and what each regime tests.

\begin{table}[h]
\centering
\small
\setlength{\tabcolsep}{4pt}
\caption{\textbf{Evaluation regimes in order of increasing difficulty.} All three regimes evaluate the same task (candidate-conditioned caller attribution); they differ in which axes of variation are held out at test time.}
\label{tab:regimes}
\begin{tabular}{llll}
\toprule
\textbf{Regime} & \textbf{Held out at test} & \textbf{Held fixed} & \textbf{What it isolates} \\
\midrule
Session-disjoint        & Recording date              & Birds, cage, array          & In-domain session noise \\
Held-out experiment     & Bird cohort + acq.\ context & No birds shared             & Identity + context shift \\
Leave-one-bird-out      & Positive caller examples    & Cage, array, cohort, dates  & Held-out vocalizer \\
\bottomrule
\end{tabular}
\end{table}

\paragraph{Metrics.}
\label{sec:metrics}
Top-1 attribution accuracy is the primary metric on every split. In leave-one-bird-out, the uniform fold mean is also per-bird recall because each test fold contains positive calls from one held-out physical bird; Appendix~\ref{sec:appendix_weighted_loo} reports the sample-weighted companion. 3D source localization is a secondary diagnostic, split into \emph{free} localization, which regresses the 3D source position, and \emph{candidate-conditioned} localization, induced from candidate scores or the selected candidate's position~\citep{evers2020locata,politis2022dcase,shimada2023starss}. Collapsing the two would conflate models that genuinely localise with models that only re-rank a known candidate set.

\paragraph{Reporting.}
\label{sec:reporting}
For learned models, session-disjoint and held-out-experiment cells report mean $\pm$ standard deviation across training seeds. Leave-one-bird-out cells average seeds within each fold and then report mean $\pm$ standard deviation across the 10 folds, so fold-level variation stays visible rather than being smoothed away by seed averaging. Deterministic baselines are reported without $\pm$. The withheld on-body reference is shown in every regime as a ceiling rather than as a competitor, because it uses the verification channel withheld from all other models.

\section{Baselines}
\label{sec:baselines}

All baselines except the on-body reference solve the same task: given an event, choose which visible bird produced the vocalization. We group them into six families by how they turn the multimodal inputs into a per-candidate score, since the benchmark question is which formulation transfers rather than which backbone fits best in-domain. Full architectures and modality usage are detailed in Appendix~\ref{app:methods}.

\paragraph{Chance and physics references.}
These rows establish how far a model can get without learned audio identity. \textbf{Random} picks one visible bird uniformly. \textbf{LoudestMic} attributes the call to the visible bird closest in 3D to the loudest microphone, with no time differences and no training. \textbf{SRP~$\pm 20$}~\citep{knapp1976gccphat} scores each candidate by how well the observed inter-microphone time differences match those expected from its location, again without voice identity or label training. \textbf{SRP reranker} layers a learned reranker on the same evidence, and \textbf{NeuralSRP}~\citep{nguyen2022salsalite} replaces the handcrafted spatial features with a learned spatial representation while staying geometry-first.

\paragraph{Slot-indexed audio classifier.}
\textbf{SELDnet}~\citep{adavanne2018seldnet} predicts a candidate slot index directly from the multichannel audio through a fixed 4-way head. Because the slot-to-bird mapping is stable within an experiment, on session-disjoint evaluation the head effectively learns per-bird audio signatures, while on leave-one-bird-out the held-out slot is absent from training by construction and the model has no mechanism to match audio against candidate state. This row represents the formulation that looks strong on familiar individuals but is structurally unable to generalise to unseen ones.

\paragraph{Candidate-scoring models.}
These score each visible bird from audio plus its candidate-side information (3D position, head direction, radio), so they can in principle identify unseen birds. \textbf{SELDnet++} extends SELDnet with full candidate conditioning and is the only row using every candidate-side modality; \textbf{SELDnet++ open} removes the slot head to force pure candidate matching. \textbf{SpatialScorer} combines microphone-pair time-difference features with per-candidate geometry and cross-experiment regularisation. \textbf{PairRank} learns a pairwise preference between candidates, \textbf{Contrastive} embeds audio and each candidate's spatial signature in a shared space and attributes by cosine similarity, and \textbf{Whisper candidate}~\citep{radford2023whisper} swaps the audio backbone for frozen Whisper features; Appendix~\ref{app:frontends} also tests frozen bioacoustic front ends (AVES, BirdNET, animal2vec) under a candidate-scoring head.

\paragraph{Localization-first and literature-style spatial-audio baselines.}
These route attribution through localization, estimating a 3D source position and picking the nearest visible bird. \textbf{AudioANN} and \textbf{RadioANN} ship in three variants each (\emph{direct}, \emph{locfirst}, \emph{hybrid}), so the effect of routing through localization is visible within a single backbone. \textbf{ACCDOA}~\citep{shimada2022multiaccdoa}, \textbf{EINV2}~\citep{cao2021einv2}, and \textbf{CST-former}~\citep{shul2024cstformer} are literature-style array-audio architectures adapted only at the attribution head (Appendix~\ref{app:budget} reruns CST-former and Whisper candidate with the 60-epoch budget used for SELDnet), and \textbf{BeamFusion} adds learned per-candidate beamforming with transformer fusion.

\paragraph{Vision-only controls and withheld reference.}
\textbf{PoseMotion} scores candidates from 3D keypoint trajectories and \textbf{VideoCandidate} from per-bird camera crops, in the spirit of visible-speaker attribution in human video~\citep{roth2020ava,tao2021someone,nguyen2025unitalk}, measuring how much of attribution is already visible from video alone before audio and geometry combine. The \textbf{withheld on-body reference} uses the on-body vibration channel that anchors the labels and is reported as a ceiling rather than as a deployable baseline. Its performance measures the information available in the withheld verification channel and serves as a consistency check for the retained events.


\section{Experiments}
\label{sec:experiments}

We evaluate all baselines under the three benchmark regimes on the same events and candidate sets. Learned rows report mean $\pm$ std across seeds; LOO entries average seeds within each fold before reporting mean $\pm$ std across folds, so fold heterogeneity is not smoothed away. The withheld on-body reference uses the verification channel and is reported only as a ceiling.

\begin{table}[h!]
\centering
\caption{\textbf{Vocal attribution and LOO modality interventions across benchmark regimes (representative subset) using V33DA.} Attribution accuracy (\%) for one representative method per family; the full 24-method matrix is in Appendix Table~\ref{tab:main_matrix_full}. The last three columns are LOO accuracy changes (points) when one input channel is removed at test time on a single seed-42 checkpoint: zeroing the audio ($\Delta$aud), shuffling candidate geometry across events ($\Delta$geo), or zeroing the radio features ($\Delta$rad). Negative values indicate degradation. \textbf{Session} tests new recordings of known birds; \textbf{Held-out BP01/BP02/BP05} rotate the test experiment; \textbf{LOO mean} averages within-experiment leave-one-bird-out folds. Learned rows report mean $\pm$ std across seeds. The withheld on-body reference is a ceiling.}
\label{tab:main_matrix}
\small
\setlength{\tabcolsep}{3pt}
\resizebox{\textwidth}{!}{%
\begin{tabular}{lccccccccc}
\toprule
 &  &  &  &  & \multicolumn{4}{c}{LOO mean} \\
\cmidrule(lr){6-9}
Method & Session & Held-out BP01 & Held-out BP02 & Held-out BP05 & Base & \cellcolor{red!10}$\Delta$aud & \cellcolor{blue!10}$\Delta$geo & \cellcolor{green!10}$\Delta$rad \\
\midrule
SRP $\pm$20 (physics)       & 51.0                    & 52.4                    & 62.2                    & 35.8                    & 53.5 $\pm$ 27.5         & \cellcolor{black!8}-- & \cellcolor{black!8}-- & \cellcolor{black!8}-- \\
\midrule
SELDnet++ (slot)            & \textbf{86.5 $\pm$ 0.4} & 39.8 $\pm$ 0.8          & 44.0 $\pm$ 1.2          & 36.9 $\pm$ 3.5          & 0.0 $\pm$ 0.0           & $+0.0$ & $+0.0$ & $+0.0$ \\
SpatialScorer               & 73.0 $\pm$ 1.2          & 56.7 $\pm$ 1.6          & \textbf{69.1 $\pm$ 1.0} & \textbf{55.0 $\pm$ 1.2} & 49.5 $\pm$ 22.0         & $-26.2$ & $-25.2$ & $+6.3$ \\
ACCDOA                      & 65.6 $\pm$ 0.2          & 58.1 $\pm$ 2.0          & 63.4 $\pm$ 1.3          & 38.6 $\pm$ 0.9          & \textbf{55.4 $\pm$ 21.8}& $-27.4$ & $-29.0$ & $+0.0$ \\
EINV2                       & 65.8 $\pm$ 0.3          & \textbf{58.2 $\pm$ 1.9} & 66.6 $\pm$ 1.4          & 38.2 $\pm$ 0.3          & 55.2 $\pm$ 22.1         & $-26.4$ & $-26.9$ & $+0.0$ \\
AudioANN-direct (slot)      & 72.1 $\pm$ 3.1          & 35.0 $\pm$ 0.4          & 42.4 $\pm$ 0.5          & 28.8 $\pm$ 12.0         & 0.0 $\pm$ 0.0           & $+0.0$ & $+0.0$ & $+0.0$ \\
\midrule
\rowcolor{black!8}
On-body ref.                & 99.2 $\pm$ 0.1          & 91.9 $\pm$ 1.0          & 98.0 $\pm$ 0.6          & 96.6 $\pm$ 0.2          & 91.3 $\pm$ 7.6          & --     & --     & --     \\
\bottomrule
\end{tabular}}
\end{table}

\paragraph{Attribution (Table~\ref{tab:main_matrix}).}
The best session model is the worst transfer model. SELDnet++ reaches 86.5\% on session-disjoint evaluation but drops to 0\% under leave-one-bird-out. SELDnet and AudioANN-direct fail in the same way. All three carry an identity-slot output head: at training time the model learns one slot per training bird, and when the test bird is unfamiliar, no slot fits.

The five rows that transfer (EINV2, SpatialScorer, ACCDOA, SRP, and AudioANN-locfirst) instead score the visible candidates directly, or attribute through localization. They include a classical physics method (SRP) and learned models, so the discriminating factor is the spatial formulation, not the choice of architecture. Their fine-grained ordering, however, is not resolved by ten LOO folds and three held-out experiments (Appendix~\ref{sec:appendix_intervention_detail}).

The withheld on-body reference stays above 91\% in every regime, indicating that the retained events are internally consistent under the withheld verification signal and leaving a substantial gap for the benchmark models. Figure~\ref{fig:overview}(g--i) shows the same contrast at the representation level. Appendix~\ref{sec:appendix_attribution_detail} gives the slot-collapse analysis, regime-level results, and sample-weighted LOO confirmations.

\paragraph{Close birds.}
Birds that sit close together leave the array a smaller timing margin. Stratifying LOO events by the distance from the vocalizer to its nearest competing candidate, ACCDOA falls from 73.0\% above 50~cm to 46.6\% below 10~cm, while random attribution changes by only 4.8 points (Appendix~\ref{app:distance}). Even below 10~cm, a fixed GCC-PHAT decoder that uses no learned identity reaches 36.6\% against 26.3\% for random attribution, so close spacing is a measurable source of difficulty rather than a loss of spatial information.

\paragraph{Localization.}
We check whether the audio-spatial rows have learned a usable source position, rather than only a candidate ranking. Table~\ref{tab:loc_main} reports a representative subset; the full 16-row table is in Appendix Table~\ref{tab:localization_matrix}. The results match the attribution picture: EINV2 and CST-former give the lowest errors in free 3D localization, and SpatialScorer and BeamFusion among candidate-conditioned methods. On LOO, SpatialScorer reaches 19.7~cm, about 7~cm below the random-candidate baseline (27.1~cm) and above the on-body reference (3.3~cm); this closes 31\% of the range between the two, comparable to the 35\% it closes in attribution accuracy (49.5\% against 27.0\% random and 91.3\% for the reference). The same gap is even tighter on held-out BP05.

\begin{table}[h!]
\centering
\caption{\textbf{Localization across benchmark regimes (representative subset) using V33DA.} Mean error in cm (lower is better). \emph{Setting}: Free = unconstrained 3D regression; Cand. = candidate-conditioned (one of the visible birds). The full table is in Appendix Table~\ref{tab:localization_matrix}. The random-candidate row picks one visible candidate uniformly at random; the shaded withheld on-body reference row is a ceiling for the candidate-conditioned setting only.}
\label{tab:loc_main}
\small
\setlength{\tabcolsep}{4pt}
\begin{tabular}{llccccc}
\toprule
Setting & Method & Session & Held-out BP01 & Held-out BP02 & Held-out BP05 & LOO mean \\
\midrule
Free  & EINV2             & \textbf{20.7 $\pm$ 0.6} & 23.9 $\pm$ 0.8          & \textbf{21.7 $\pm$ 0.4} & 27.1 $\pm$ 0.7          & 25.0 $\pm$ 9.6 \\
      & CST-former        & 21.8 $\pm$ 0.4          & \textbf{22.8 $\pm$ 0.7} & 25.8 $\pm$ 0.4          & \textbf{26.9 $\pm$ 0.5} & \textbf{23.2 $\pm$ 5.0} \\
\midrule
Cand. & Random candidate  & 27.5                    & 26.9                    & 27.1                    & 28.4                    & 27.1 $\pm$ 1.3 \\
      & SpatialScorer     & 18.2 $\pm$ 0.1          & \textbf{17.4 $\pm$ 0.1} & \textbf{19.2 $\pm$ 0.2} & 24.2 $\pm$ 0.3          & \textbf{19.7 $\pm$ 5.5} \\
      & BeamFusion        & \textbf{11.7 $\pm$ 0.7} & 17.1 $\pm$ 1.6          & 17.1 $\pm$ 1.1          & \textbf{20.5 $\pm$ 0.6} & 23.7 $\pm$ 6.7 \\
\rowcolor{black!8}
Ref. & On-body ref.      & 0.3 $\pm$ 0.0           & 2.9 $\pm$ 0.4           & 0.8 $\pm$ 0.2           & 1.4 $\pm$ 0.1           & 3.3 $\pm$ 3.1 \\
\bottomrule
\end{tabular}
\end{table}

\paragraph{Modality interventions.}
\label{sec:diagnostics}
The last three columns of Table~\ref{tab:main_matrix} apply three test-time interventions to the transfer rows, each isolating one input channel: zeroing the audio, shuffling candidate geometry across events, and zeroing the radio features.

For ACCDOA and EINV2 the pattern holds on held-out BP01, held-out BP02, and LOO: removing audio costs 25--30 points and shuffling geometry costs 27--33 points, whereas zeroing the radio representation changes accuracy by less than a point, so transfer is carried by clean audio together with correct candidate geometry, with radio largely redundant once geometry is known. On held-out BP05, where every method is weakest, these audio and geometry effects shrink to roughly 7--12 points. SpatialScorer is a formulation-dependent exception on BP05, where zeroing radio costs 14.8 points. Position-noise robustness and cross-model significance tests (Appendix~\ref{sec:appendix_intervention_detail}) split the rows along the same geometry-using vs geometry-agnostic line, confirming the central lesson: V33DA rewards usable spatial structure, not generic multimodality.


\paragraph{What V33DA reveals.}
\label{sec:discussion}
High session-split accuracy does not imply transferable attribution. The matrix separates methods that exploit familiar individuals from methods that preserve usable spatial structure for unseen individuals, and four observations make this concrete. Familiar-voice recognition is effective when every test bird is familiar: SELDnet++ reaches 86.5\% session-disjoint. The same formulation does not reach held-out vocalizers: with an identical encoder and training, AudioANN reaches 0.0\% LOO when it attributes directly and 53.0\% when it attributes through localization. Candidate conditioning alone is not sufficient: SELDnet++ open, PairRank, Whisper candidate, and CST-former all remain below the 27.0\% random reference on LOO. And the rows that transfer rely on the array: removing microphones lowers SpatialScorer from 51.2\% to 26.0\% LOO, and permuting microphone channels lowers ACCDOA from 64.3\% to 38.8\% session-disjoint (Appendix~\ref{app:mics}). The strongest transfer rows share a formulation, not a backbone: they score visible candidates directly, reason over spatial geometry, or route attribution through localization. Multi-individual attribution should therefore be evaluated as transfer rather than held-out-date accuracy on familiar individuals, identity shift should be separated from broader experiment shift when possible, and a model's actual modality use should be checked directly rather than inferred from headline accuracy.

The on-body channel serves here to construct trustworthy labels, not as a permanent recording requirement. Attribution from room sensors is most needed where voice identification of familiar individuals breaks down: juvenile subsong changes rapidly and carries little individual signature, rare vocalizations such as copulation calls may be too sparse to learn, and new birds enter a cohort without labeled calls. Acoustic clustering does not remove this need: even with segmented events, the true number of birds, and an encoder trained on the same cohort, $k$-means reaches NMI 0.61 and ARI 0.58, and the clusters must still be mapped to the tracked candidates (Appendix~\ref{app:frontends}).


\section{Limitations}
\label{sec:limitations}

V33DA has three scope boundaries. First, all data come from a single sensing installation with fixed cage, microphone, camera, and 4-antenna geometry, covering 3 experiments, 10 birds, and 33{,}625 verified vocalizations; cross-installation transfer is outside the scope of our regimes. Second, the per-bird radio transmitters lack a frequency-stabilizing quartz, so the carrier drifts faster than the released analysis window; the 12-dimensional radio summary captures only the standard deviation of the carrier, not its fast temporal structure, and prior work \citep{ruttimann2024birdpark} has shown these sub-framerate dynamics encode fast movements such as wingflaps. The apparent redundancy of radio for localization (Table~\ref{tab:main_matrix}) likely reflects this feature summary, not the modality itself. Third, the V33DA benchmark covers only verified single-vocalizer events with stable candidate state; multi-caller attribution, source separation, vocal activity detection, and longer-context tasks are left to V33DA++, which is released as an unevaluated companion (§\ref{sec:dataset}).

\section{Conclusion}
\label{sec:conclusion}

We introduced V33DA, a physically verified multimodal benchmark for spatial vocal attribution in social zebra finches. Across session-disjoint, held-out-experiment, and leave-one-bird-out evaluation, V33DA shows that high in-domain accuracy can come from bird-identity shortcuts, while transfer to unseen individuals requires spatial reasoning. Slot-indexed audio classifiers look strong in-domain but collapse under unseen birds; candidate-aware spatial and localization-first methods transfer, but remain well below the withheld on-body reference, which measures the ceiling available when the withheld verification channel is observed. Beyond caller attribution, the same synchronized sensor state supports audio, video, pose, and radio modeling.

\begin{ack}
This work was supported by the Swiss National Science Foundation (grant 31003A\_182638) and the NCCR Evolving Language, Swiss National Science Foundation Agreement No.~51NF40\_180888, and by a collaborative project with the Swiss Data Science Center (SDSC). The funders had no role in study design, data collection, analysis, or manuscript preparation. The authors declare no competing interests.
\end{ack}

\bibliographystyle{plainnat}
\bibliography{references}


\section*{Ethics, Licensing, and Release}

All animal procedures were approved by the Cantonal Veterinary Office of Zurich under license ZH148/2020 and comply with Swiss and EU animal welfare regulations. The birds were housed in enriched environments with \emph{ad libitum} food and water.

The benchmark can support research on animal communication, multimodal learning, and automated behavioral analysis under controlled, physically verified supervision. We do not currently foresee material societal risks from releasing V33DA itself. A remote concern is that attribution or localization systems benchmarked in this setting could eventually be repurposed for surveillance-style monitoring in other contexts, but that possibility is far removed from the present dataset, task, and intended use.

The dataset is released under a CC-BY-4.0 license at \url{https://huggingface.co/datasets/songbirdini/v33da}, with the unevaluated V33DA++ companion release at \url{https://huggingface.co/datasets/songbirdini/v33da_pp}. The accompanying code and baseline implementations are available at \url{https://github.com/marisbasha/v33da}. External software and pretrained components used in preprocessing and comparison, including WhisperSeg/Whisper, YOLOX/YOLO-Pose, and OC-SORT, are credited in the paper and remain subject to their original licenses and terms of use. The dataset contains no personally identifiable information.

\paragraph{Artifact availability.}
The submission packet includes the public V33DA dataset URL, the executable code URL, and a Croissant JSON file describing the main V33DA benchmark artifact. V33DA++ is linked as an unevaluated companion release of overlap and padded-context files.

\appendix
\let\oldsubsection\subsection
\renewcommand{\subsection}{\FloatBarrier\oldsubsection}
\section{Task Coverage of Related Resources}
\label{app:taskcoverage}

\begin{minipage}{\columnwidth}
\centering
\scriptsize
\setlength{\tabcolsep}{3pt}
\captionof{table}{\textbf{Task coverage of V33DA versus related resources.} A checkmark indicates that the released data directly support the task. In the V33DA row, a star ($\star$) marks tasks evaluated in this paper, and a dagger ($\dagger$) marks tasks for which the unevaluated V33DA++ companion release provides data.}
\label{tab:task_coverage}
\resizebox{\columnwidth}{!}{%
\begin{tabular}{lccccccccccc}
\toprule
 & \multicolumn{4}{c}{\textit{Audio tasks}} & \multicolumn{2}{c}{\textit{Multimodal}} & \multicolumn{3}{c}{\textit{Video tasks}} & \textit{Radio} & \\
\cmidrule(lr){2-5}\cmidrule(lr){6-7}\cmidrule(lr){8-10}\cmidrule(lr){11-11}
 & \textbf{SSL} & \textbf{Sep} & \textbf{VAD} & \textbf{ID} & \textbf{CA} & \textbf{ASD} & \textbf{KP} & \textbf{Pose} & \textbf{Track} & \textbf{Rad} & \textbf{Transfer} \\
\midrule
Anisimov+, Gill+ \citeyearpar{anisimov2014reconstruction,gill2016backpack}       & -- & \checkmark & -- & -- & \checkmark & -- & -- & -- & -- & -- & -- \\
Rookognise \citeyearpar{music2022rookognise}                                     & -- & -- & \checkmark & \checkmark & \checkmark & -- & -- & -- & -- & -- & -- \\
MeerKAT \citeyearpar{stoeger2024meerkat}                                         & -- & -- & \checkmark & -- & -- & -- & -- & -- & -- & -- & -- \\
Prat+ \citeyearpar{prat2017annotated}                                            & -- & -- & -- & -- & \checkmark & -- & -- & -- & -- & -- & -- \\
HyVL \citeyearpar{hyvl2023}                                                      & \checkmark & \checkmark & \checkmark & -- & \checkmark & -- & -- & -- & \checkmark & -- & -- \\
VCL \citeyearpar{vcl2024}                                                        & \checkmark & \checkmark & \checkmark & -- & \checkmark & -- & -- & -- & -- & -- & -- \\
Elie+, Phaniraj+ \citeyearpar{elie2018zebra,zuercher2023caller}                  & -- & -- & -- & \checkmark & -- & -- & -- & -- & -- & -- & -- \\
VoxCeleb \citeyearpar{nagrani2020voxceleb}                                       & -- & -- & -- & \checkmark & -- & -- & -- & -- & -- & -- & \checkmark \\
AVA-ActSp., UniTalk \citeyearpar{roth2020ava,nguyen2025unitalk}                  & -- & -- & -- & -- & -- & \checkmark & -- & -- & -- & -- & \checkmark \\
AVA-AVD \citeyearpar{xu2022avaavd}                                               & -- & -- & -- & -- & \checkmark & \checkmark & -- & -- & -- & -- & \checkmark \\
Ego4D \citeyearpar{grauman2022ego4d}                                             & -- & -- & -- & -- & -- & \checkmark & -- & -- & -- & -- & \checkmark \\
AMI \citeyearpar{carletta2005ami}                                                & -- & \checkmark & \checkmark & \checkmark & \checkmark & \checkmark & -- & -- & -- & -- & -- \\
DIHARD II \citeyearpar{ryant2019dihard}                                          & -- & -- & \checkmark & -- & -- & -- & -- & -- & -- & -- & \checkmark \\
VoxConverse \citeyearpar{chung2020voxconverse}                                   & -- & -- & \checkmark & -- & \checkmark & \checkmark & -- & -- & -- & -- & \checkmark \\
STARSS23 \citeyearpar{shimada2023starss}                                         & \checkmark & -- & \checkmark & -- & -- & -- & -- & -- & -- & -- & -- \\
DeepLabCut, SLEAP, idTracker \citeyearpar{mathis2018deeplabcut,pereira2022sleap,romeroferrero2019idtracker} & -- & -- & -- & -- & -- & -- & \checkmark & -- & \checkmark & -- & -- \\
3D-POP, 3D-MuPPET \citeyearpar{naik2023threedpop,naik2024threedmuppet}           & -- & -- & -- & -- & -- & -- & \checkmark & \checkmark & \checkmark & -- & -- \\
CalMS21 \citeyearpar{sun2021calms21}                                             & -- & -- & -- & -- & -- & -- & \checkmark & -- & \checkmark & -- & -- \\
\midrule
\rowcolor{black!5}
\textbf{V33DA (ours)} & \textbf{\checkmark$\star$} & \textbf{\checkmark$\dagger$} & \textbf{\checkmark$\dagger$} & \textbf{\checkmark} & \textbf{\checkmark$\star$} & \textbf{\checkmark$\dagger$} & \textbf{\checkmark} & \textbf{\checkmark} & \textbf{\checkmark$\dagger$} & \textbf{\checkmark} & \textbf{\checkmark$\star$} \\
\bottomrule
\end{tabular}%
}
\end{minipage}


\section{Dataset Construction Pipeline}
\label{app:pipeline}

This appendix documents the full curation pipeline that converts raw BirdPark recordings~\citep{ruttimann2024birdpark} into the released V33DA events. We describe raw acquisition, candidate proposal from on-body sensing, vision-based state reconstruction, identity assignment, 3D triangulation, manual verification, and post-review quality filtering. For each stage we specify the inputs, the processing steps, and the criteria by which events are retained or removed, so that the release is auditable end to end.

\subsection{Raw Data Acquisition}

Each recording session produces a 7-minute H5 file containing:
\begin{itemize}[nosep]
    \item \texttt{daqSignals}: $(7, N)$ float32: 7 DAQ channels (5 cage microphones + 2 nest microphones) at 24{,}414~Hz. BP05 additionally logs a room reference microphone, giving $(8, N)$; only the 5 cage microphones are used in the benchmark.
    \item \texttt{radioSignals}: $(N_\text{birds}, N)$ float32: FM-demodulated instantaneous transmitter frequency from body-mounted transmitters at 24{,}414~Hz.
    \item \texttt{radioFrameSignals}: 21 per-bird per-frame features (tracking frequency, master- and rotated-vector phase angles, per-antenna compensation phases, and baseband signal powers and variances) tracked at video frame rate (47.68~Hz).
    \item Synchronized composite video (2976$\times$2016, 47.68~fps) combining top, back, and side camera views.
\end{itemize}

Table~\ref{tab:raw_data} summarizes the raw recording volumes. In total, approximately 74~hours of continuous multi-modal recordings were processed to extract the 33{,}625 vocalization events in the released dataset.

\begin{table}[h]
\centering
\small
\caption{Raw data volumes per experiment.}
\label{tab:raw_data}
\begin{tabular}{lccc}
\toprule
Experiment & Days & H5 Files & Recording Hours \\
\midrule
juvExpBP01 & 10 & $\sim$400 & $\sim$46 \\
juvExpBP02 & 3 & $\sim$120 & $\sim$14 \\
juvExpBP05 & 3 & $\sim$120 & $\sim$14 \\
\bottomrule
\end{tabular}
\end{table}

\subsection{Vocalization Segmentation}
\label{app:segmentation}

Vocalization candidates are proposed from the body-mounted accelerometer-derived signals using a multichannel WhisperSeg model~\citep{gu2024whisperseg,radford2023whisper}, a Whisper-large variant fine-tuned for bioacoustic segmentation. WhisperSeg is used only to propose candidate windows; it does not define final benchmark labels. For each bird, the model receives three input channels:
\begin{enumerate}[nosep]
    \item the target bird's radio/accelerometer signal,
    \item the sum of the other birds' radio/accelerometer signals,
    \item the sum of the cage-microphone signals.
\end{enumerate}
Each channel is peak-normalized to $[-1, 1]$, resampled to 16~kHz, converted to an 80-bin mel spectrogram, concatenated to form a 240-bin representation, and interleaved back to 80 bins before being processed by the Whisper encoder-decoder. The model predicts onset and offset times for each candidate vocalization. The summed cage-microphone channel is used only as a shared timing and event-presence cue; final caller identity is established downstream by comparing the target bird's on-body vibration signature to those of the other birds, not by localizing the call from room acoustics alone. Candidate windows are discarded if the event is not visible in the microphone channels, if the on-body evidence is ambiguous, or if overlapping vocalizations are present. Upstream misses reduce coverage of the released event set but do not corrupt the labels of retained events: V33DA contains only candidate windows that survive manual verification and post-review filtering.

\subsection{2D Detection and Keypoint Estimation}
\label{app:detection}

\paragraph{Bird detection and keypoint estimation.}
YOLOXPose~\citep{ge2021yolox,maji2022yolo} performs joint bird detection and five-keypoint estimation in each camera view. The model runs full-frame inference at $640 \times 640$ input resolution with a detection score threshold of 0.8 and NMS threshold of 0.7, producing per-bird bounding boxes and five 2D keypoints: beak tip, head center, backpack center, tail base, and tail tip. Separate YOLOXPose checkpoints (Tiny variant) are trained for each camera view (top, back, side) on a pool of ${\sim}$4{,}400 manually annotated frames collected across multiple BirdPark recording sessions (including but not limited to the three experiments in V33DA). Each annotated frame includes five keypoints per visible bird with visibility flags and color-coded identity labels. Because camera placement varies between experiments, we fine-tune the detector per-experiment using the subset of annotations from that experiment's camera configuration.

\paragraph{2D tracking.}
OC-SORT~\citep{cao2023ocsort} maintains consistent track identities across frames within each camera view independently, with a history buffer of 500 frames.

\subsection{Identity Assignment via Backpack Color Detection}
\label{app:identity}

Each bird wears a uniquely colored backpack. Identity is assigned within each vocalization window from the tracked top-view detections using the color stage in the BirdPark tracking pipeline.

\paragraph{Stage 1: Frame selection.}
Color inference is run only on the top view and only on selected frames within the vocalization window. The tracker first identifies informative frames around track starts, ends, and long gaps, then adds periodic samples between those events. Within each sampling window, the pipeline picks the frame with the best geometric support, favoring frames with visible backpack keypoints and good 2D bird separation.

\paragraph{Stage 2: Color detection and track assignment.}
On each selected frame, a K-NN color classifier assigns backpack color from a pre-built gallery of labeled patches. The classifier extracts a $24 \times 24$ patch in CIELAB color space centered on the backpack keypoint, flattens it, and queries the $K{=}11$ nearest neighbors from the gallery (approximately 4{,}700 labeled patches per experiment) using L2 distance. The majority vote determines the predicted color. A 1-second temporal smoothing window (47~frames at the native frame rate) and a minimum run length of 3~seconds suppress spurious predictions. Per-frame assignments with confidence below 0.5 are discarded. The result is a set of per-frame color observations for each top-view track.

\paragraph{Stage 3: Temporal stabilization.}
Per-frame color observations are converted into track-level identities with confidence pre-filtering and hysteresis segmentation. The dominant color segment for each track is retained, and the resulting color label is mapped to bird identity using the experiment metadata. Final dataset release does not depend on this automatic step alone: all released events are manually reviewed, and candidates with uncertain or inconsistent identities are excluded from release.

\subsection{Pipeline Component Evaluation}
\label{app:pipeline_eval}

\paragraph{Color classification accuracy.}
Table~\ref{tab:knn_color_eval} reports the K-NN backpack color classification accuracy on held-out annotated frames for the three V33DA experiments. Per-experiment galleries are built exclusively from each experiment's own annotated frames to prevent cross-experiment color confusion. Patches with occluded backpacks (COCO visibility $<$ 2) are excluded from both the gallery and evaluation set. Each clean patch is augmented $48\times$ (4 rotations $\times$ 2 flips $\times$ 3 brightness offsets $\times$ 2 blur levels) before gallery construction. Outlier removal via leave-one-out K-NN consensus further filters mislabeled or ambiguous patches.

\begin{table}[h]
\centering
\small
\caption{K-NN backpack color classification accuracy per experiment.}
\label{tab:knn_color_eval}
\begin{tabular}{lccc}
\toprule
Experiment & Clean patches & After 48$\times$ aug & Accuracy \\
\midrule
juvExpBP01 & ${\sim}$2{,}000 & ${\sim}$96{,}000 & 98.1\% \\
juvExpBP02 & ${\sim}$1{,}050 & ${\sim}$50{,}400 & 95.2\% \\
juvExpBP05 & ${\sim}$1{,}040 & ${\sim}$49{,}920 & 97.3\% \\
\midrule
\textbf{Average} & & & \textbf{96.9\%} \\
\bottomrule
\end{tabular}
\end{table}

\paragraph{Ablation: gallery construction.}
Averaging across 15 BirdPark experiments (including but not limited to the three V33DA experiments), color accuracy improves from 88.9\% (shared gallery, baseline) to 91.2\% (per-experiment galleries), 93.0\% (+ outlier removal), 94.8\% (+ occlusion filtering), and 96.4\% (+ 48$\times$ augmentation).

\paragraph{Keypoint quality.}
Keypoint quality is assessed indirectly through 3D-to-2D reprojection error after triangulation. Median reprojection errors on released samples are 7.9\,px (BP01), 2.5\,px (BP02), and 12.3\,px (BP05), corresponding to roughly 1--2\,mm spatial precision at cage scale. Samples with reprojection error $>$40\,px are excluded (see Section~\ref{app:filtering}).

\subsection{Cross-View Assignment and 3D Triangulation}
\label{app:triangulation}

\paragraph{Camera calibration.}
Per-experiment camera calibrations are computed from manually annotated 3D reference points on the cage structure (cage corners, perch endpoints, microphone positions), following standard multi-view geometry practice~\citep{hartley2003multiple}. We use \texttt{cv2.calibrateCamera()} to estimate intrinsic matrices $\mathbf{K}$, distortion coefficients, and extrinsic parameters (rotation $\mathbf{R}$, translation $\mathbf{t}$) for each view. Calibration reprojection error on reference points is 2.5--3.7~px across experiments.

\paragraph{Cross-view matching.}
To establish correspondence between 2D detections across the top and back camera views, we compute the Sampson distance between each pair of detections using the fundamental matrix derived from the calibrations~\citep{hartley2003multiple}. For each vocalization event, we enumerate all possible top--back detection pairs and select the assignment that minimizes total Sampson distance while maintaining geometric consistency.

\paragraph{Triangulation.}
Matched 2D keypoint pairs are triangulated using the Direct Linear Transform (DLT) method~\citep{hartley2003multiple}. Given a 2D point $\mathbf{x}_1$ in view 1 and $\mathbf{x}_2$ in view 2, we construct the system:
\begin{equation}
\mathbf{A}\mathbf{X} = \mathbf{0}, \quad \mathbf{A} = \begin{bmatrix} x_1 \mathbf{p}_1^{3\top} - \mathbf{p}_1^{1\top} \\ y_1 \mathbf{p}_1^{3\top} - \mathbf{p}_1^{2\top} \\ x_2 \mathbf{p}_2^{3\top} - \mathbf{p}_2^{1\top} \\ y_2 \mathbf{p}_2^{3\top} - \mathbf{p}_2^{2\top} \end{bmatrix}
\end{equation}
where $\mathbf{p}_i^{j\top}$ is the $j$-th row of the projection matrix $\mathbf{P}_i = \mathbf{K}_i [\mathbf{R}_i | \mathbf{t}_i]$. The 3D point $\mathbf{X}$ is recovered via SVD. Triangulation does not use lens distortion (\texttt{distCoeffs=None}) as the calibration reference points span a limited field of view where distortion effects are minimal.

\paragraph{Coordinate system.}
All 3D coordinates are expressed in Z-up convention (millimeters): $X$ = left--right, $Y$ = front--back, $Z$ = up--down. The coordinate frame is defined by the camera calibration reference points.

\subsection{Human Verification}
\label{app:human}

\begin{center}
\vspace{-0.6em}
\includegraphics[width=0.96\linewidth]{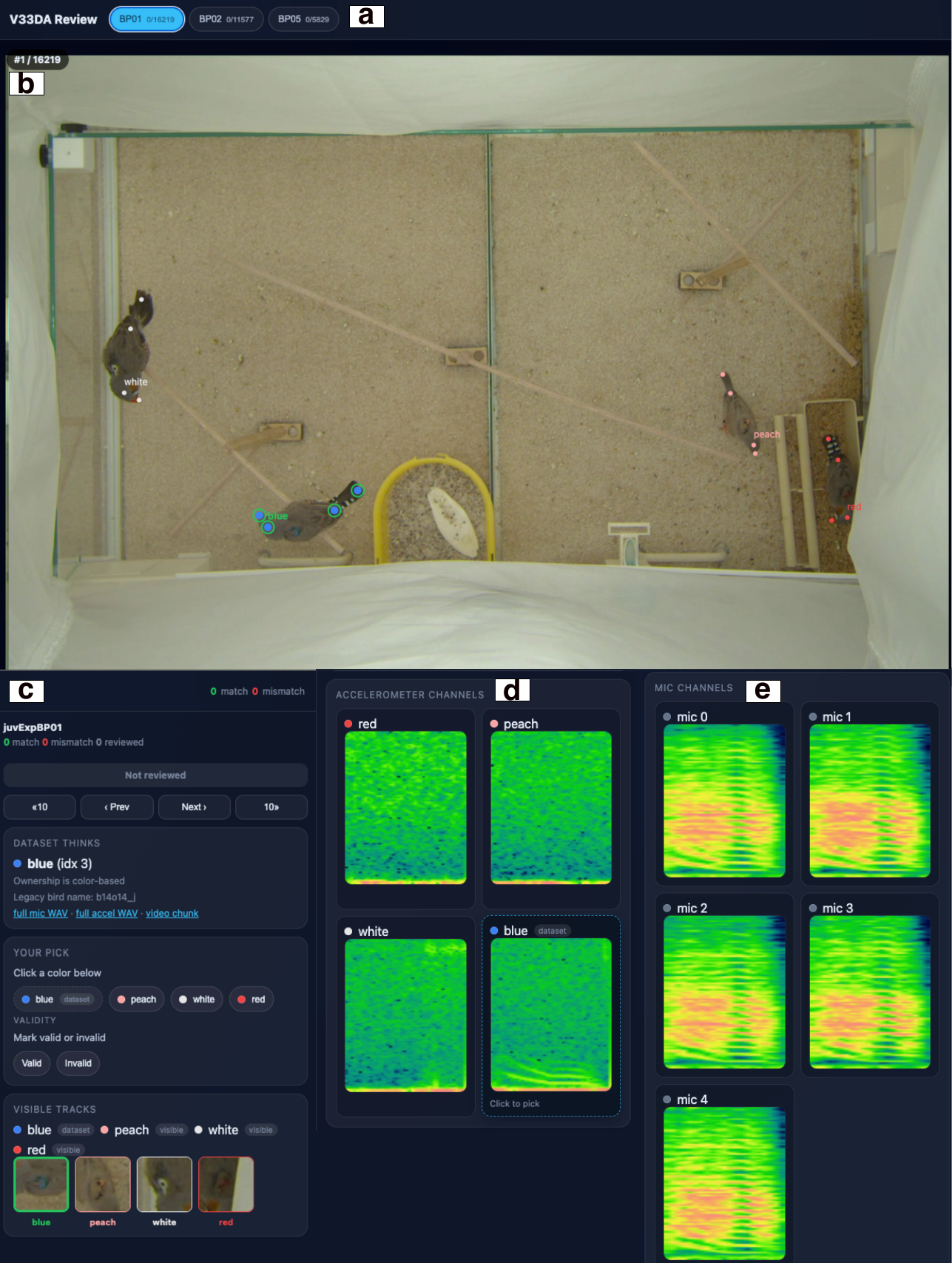}
\vspace{-0.35em}
\captionof{figure}{\textbf{Verification interface used during dataset curation.} \textbf{(a)} Experiment selector. \textbf{(b)} Synchronized video frame with overlaid keypoints and color-coded identities; the selected vocalizer is highlighted by a green ring. \textbf{(c)} Reviewer controls for navigating events, selecting the vocalizer identity, and assigning the review verdict; the crop strip shows the currently visible tracks. \textbf{(d)} Per-bird accelerometer spectrograms; the selected bird shows the vocalization-associated vibration signature. \textbf{(e)} Five-channel microphone spectrograms aligned to the same event.}
\label{fig:review_platform}
\vspace{-0.4em}
\end{center}

All candidate windows are manually reviewed by a trained annotator using the interface in Figure~\ref{fig:review_platform}. For each candidate, the reviewer inspects:
\begin{enumerate}[nosep]
    \item \textbf{Video and keypoints}: to confirm that visible birds are correctly tracked and assigned to the correct backpack identity;
    \item \textbf{Microphone audio}: to confirm that a vocalization is present at the expected time;
    \item \textbf{On-body sensing and radio telemetry}: to confirm that the labeled bird shows the characteristic thoracic-vibration signature while the event is simultaneously visible in the room microphones.
\end{enumerate}

Candidates are excluded if the automatic pipeline introduces identity swaps, missed detections, triangulation failures, ambiguous on-body evidence, or overlap with another bird's vocalization. The review stage therefore functions as a multimodal consistency check over synchronized evidence rather than as a subjective labeling step. Retained events have confirmed (a) vocalizer identity, (b) keypoint-to-bird assignment for all visible birds, and (c) temporal consistency across video, microphone audio, and on-body sensing.

\subsection{Quality Filtering and Validation}
\label{app:filtering}

After manual review, valid candidates pass through a deterministic quality filter before entering the release:

\begin{enumerate}[nosep]
    \item \textbf{Temporal alignment.} Audio, video, radio, and accelerometer streams must agree within the $\pm$10~ms event extraction window.
    \item \textbf{3D plausibility.} The triangulated vocalizer position must lie within the cage bounds ($X \in [-91, 1155]$, $Y \in [-1564, -668]$, $Z \in [236, 1530]$~mm).
    \item \textbf{Reprojection error.} The vocalizer's 3D backpack keypoint is reprojected into the top and back views using \texttt{cv2.projectPoints()}. Candidates with reprojection error above 40~px in either view are discarded. Median reprojection errors on released events are 7.9~px (BP01), 2.5~px (BP02), and 12.3~px (BP05); the 40~px threshold corresponds conservatively to about 1--2~cm of worst-case spatial error.
    \item \textbf{2D displacement.} The maximum frame-to-frame beak displacement must not exceed 40~px in either top or back view, ensuring limited motion during the short event.
    \item \textbf{Identity consistency.} The color-based identity assignment must remain consistent across the event window.
\end{enumerate}

Of the 39{,}329 reviewed candidates, 1{,}768 are rejected during review and 37{,}561 are marked valid. The geometric and overlap filters then remove 152 additional candidates for excessive reprojection error and 3{,}784 for residual temporal overlap, yielding 33{,}625 released events.

\medskip
\noindent
\begin{minipage}{\columnwidth}
\centering
\scriptsize
\captionof{table}{\textbf{Dataset summary and release audit.} Left: released events by experiment. Right: reviewed = valid + invalid; valid is then partitioned into reprojection-filtered, overlap-filtered, and released samples.}
\label{tab:dataset}
\begin{minipage}[t]{0.48\columnwidth}
\centering
\scriptsize
\setlength{\tabcolsep}{3pt}
\textbf{Dataset summary}\\[-0.15em]
\begin{tabular}{lcccc}
\toprule
Experiment & Birds & Samples & Days & Dates \\
\midrule
BP01 & 4 & 16{,}219 & 10 & Jul--Aug '21 \\
BP02 & 3 & 11{,}577 & 3 & May '22 \\
BP05 & 3 & 5{,}829 & 3 & Jul '23 \\
\midrule
\textbf{Total} & \textbf{10} & \textbf{33{,}625} & \textbf{16} & \\
\bottomrule
\end{tabular}
\end{minipage}\hfill
\begin{minipage}[t]{0.48\columnwidth}
\centering
\scriptsize
\setlength{\tabcolsep}{4pt}
\textbf{Release audit}\\[-0.3em]
\setlength{\tabcolsep}{3pt}
\begin{tabular}{lcc}
\toprule
Stage & Count & Share \\
\midrule
Reviewed & 39{,}329 & 100.0\% \\
Valid & 37{,}561 & 95.5\% \\
Invalid & 1{,}768 & 4.5\% \\
\midrule
Reproj.\ filtered & 152 & 0.4\% \\
Overlap filtered & 3{,}784 & 10.1\% \\
Released & 33{,}625 & 89.5\% \\
\bottomrule
\end{tabular}
\end{minipage}
\end{minipage}

\subsection{Post-filtering Statistics}

\begin{table}[h]
\centering
\small
\caption{Quality metrics after filtering.}
\begin{tabular}{lccc}
\toprule
Metric & BP01 & BP02 & BP05 \\
\midrule
Samples & 16{,}219 & 11{,}577 & 5{,}829 \\
Reproj error (median, px) & 7.9 & 2.5 & 12.3 \\
3D displacement (median, mm) & 2.9 & 2.5 & 2.5 \\
2D disp top (median, px) & 4.0 & 4.3 & 4.6 \\
2D disp back (median, px) & 4.4 & 3.5 & 3.6 \\
\bottomrule
\end{tabular}
\end{table}

\subsection{Audio Extraction}

For each vocalization event, we extract the full audio from onset to offset plus a $\pm$10~ms padding before and after from the 5 cage microphones (DAQ channels 0--4). The audio is stored as 5-channel WAV files (float32, 24{,}414~Hz), with clip durations ranging from 40--784~ms (mean: 150~ms, median: 125~ms). Nest microphones (channels 5--6) are excluded as they record inside enclosed nest boxes and do not carry useful spatial information for attribution.

\subsection{Radio Telemetry Extraction}

The 21 per-bird per-frame radio signals are extracted for the video frames within the vocalization window. Following BirdRadio's definitions~\citep{ruttimann2024birdpark}, the signals are: PLL tracking frequency (\texttt{frq}); master-vector phase angle $\theta_i$ (\texttt{thetaS}) and per-antenna rotated-vector angles $\alpha_i^a$ (\texttt{thetaA--D}); per-antenna compensation phases $\phi_i^a$ (\texttt{phiA--D}) and their sum (\texttt{phiM}); and the per-antenna baseband signal powers (\texttt{powRA--RD}, \texttt{powRM}) and variances (\texttt{powNA--ND}, \texttt{powNM}), all in dBFS. BirdRadio does not compute an angle of arrival. For the 11 samples where radio frame counts fell 1 frame short of the video frame count, the last radio frame is repeated to maintain alignment.

\subsection{Accelerometer Extraction}

The raw FM-demodulated radio signal per bird is extracted at audio sample rate (24{,}414~Hz) for the same $\pm$10~ms window as the audio. This signal is the physical basis for ground-truth labeling: when a bird vocalizes, thoracic vibrations modulate the FM carrier frequency, producing a distinctive waveform in the demodulated signal. The accelerometer column is provided for label verification and is \textbf{withheld from benchmark inputs}.

\subsection{Dataset Characterization}
\label{app:characterization}

\paragraph{Per-bird vocalization counts.}
Table~\ref{tab:perbird} reports the number of vocalizations per bird in each experiment and session-disjoint split. Class imbalance is substantial: within BP01, Bird~3 still contributes the largest share of samples, while within BP05 one bird contributes only 56 events in the entire release. This imbalance is one reason we report per-fold and per-experiment breakdowns rather than relying only on pooled means.

\begin{table*}[t]
\centering
\begin{minipage}[t]{0.48\textwidth}
\centering
\small
\captionof{table}{Per-bird vocalization counts by experiment and session-disjoint split. Percentages indicate within-experiment fractions.}
\label{tab:perbird}
\resizebox{\linewidth}{!}{%
\begin{tabular}{llrrr}
\toprule
Experiment & Bird & Train & Val & Test \\
\midrule
\multirow{4}{*}{BP01} & Bird 0 & 2{,}827 (21.9\%) & 218 (21.6\%) & 918 (39.4\%) \\
    & Bird 1 & 1{,}694 (13.2\%) & 197 (19.5\%) & 386 (16.6\%) \\
    & Bird 2 & 3{,}530 (27.4\%) & 248 (24.6\%) & 466 (20.0\%) \\
    & Bird 3 & 4{,}830 (37.5\%) & 345 (34.2\%) & 560 (24.0\%) \\
\midrule
\multirow{3}{*}{BP02} & Bird 0 & 726 (16.3\%) & 844 (20.3\%) & 495 (16.7\%) \\
    & Bird 1 & 1{,}870 (42.0\%) & 1{,}540 (37.1\%) & 1{,}179 (39.7\%) \\
    & Bird 2 & 1{,}859 (41.7\%) & 1{,}771 (42.6\%) & 1{,}293 (43.6\%) \\
\midrule
\multirow{3}{*}{BP05} & Bird 0 & 753 (37.1\%) & 697 (40.9\%) & 690 (33.0\%) \\
    & Bird 2 & 21 (1.0\%) & 26 (1.5\%) & 9 (0.4\%) \\
    & Bird 3 & 1{,}255 (61.9\%) & 983 (57.6\%) & 1{,}395 (66.6\%) \\
\bottomrule
\end{tabular}
}
\end{minipage}\hfill
\begin{minipage}[t]{0.48\textwidth}
\centering
\small
\captionof{table}{Leave-one-bird-out fold sizes for all experiments.}
\label{tab:loo_sizes}
\resizebox{\linewidth}{!}{%
\begin{tabular}{llrrrr}
\toprule
Experiment & Held-out bird & Train & Val & Test & Test \% \\
\midrule
\multirow{4}{*}{BP01} & Bird 0 & 10{,}961 & 1{,}295 & 3{,}963 & 24.4\% \\
 & Bird 1 & 12{,}535 & 1{,}407 & 2{,}277 & 14.0\% \\
 & Bird 2 & 10{,}763 & 1{,}212 & 4{,}244 & 26.2\% \\
 & Bird 3 & 9{,}418 & 1{,}066 & 5{,}735 & 35.4\% \\
\midrule
\multirow{3}{*}{BP02} & Bird 0 & 8{,}569 & 943 & 2{,}065 & 17.8\% \\
 & Bird 1 & 6{,}274 & 714 & 4{,}589 & 39.6\% \\
 & Bird 2 & 5{,}989 & 665 & 4{,}923 & 42.5\% \\
\midrule
\multirow{3}{*}{BP05} & Bird 0 & 3{,}329 & 360 & 2{,}140 & 36.7\% \\
 & Bird 2 & 5{,}201 & 572 & 56 & 1.0\% \\
 & Bird 3 & 1{,}974 & 222 & 3{,}633 & 62.3\% \\
\bottomrule
\end{tabular}
}
\end{minipage}
\end{table*}

\paragraph{Leave-one-bird-out fold sizes.}
Table~\ref{tab:loo_sizes} provides per-fold sample counts for all three experiments and all released birds.

\paragraph{Inter-bird spatial separation.}
Table~\ref{tab:pairwise} reports per-pair separation statistics for all three experiments. In BP01, Birds~0 and~1 are frequently co-located (25.4\% within 100~mm), as are Birds~2 and~3 (17.5\%). BP02 has wider overall separation, with the closest pair (Birds~1--2) still averaging 346~mm. BP05 contains one comparatively ambiguous pair, Birds~2--3 (11.7\% within 100~mm), while the other BP05 pairs remain substantially farther apart. This co-occurrence structure directly explains the per-fold SRP variance in the LOO evaluation.

\begin{table*}[t]
\centering
\begin{minipage}[t]{0.48\textwidth}
\centering
\small
\captionof{table}{Pairwise inter-bird distances for all experiments.}
\label{tab:pairwise}
\resizebox{\linewidth}{!}{%
\begin{tabular}{llccc}
\toprule
Experiment & Bird pair & Mean (mm) & Median (mm) & \% $<$ 100\,mm \\
\midrule
\multirow{6}{*}{BP01} & 0--1 & 228 & 199 & 25.4\% \\
 & 0--2 & 407 & 411 & 0.6\% \\
 & 0--3 & 395 & 406 & 0.4\% \\
 & 1--2 & 415 & 424 & 0.4\% \\
 & 1--3 & 410 & 412 & 0.3\% \\
 & 2--3 & 254 & 236 & 17.5\% \\
\midrule
\multirow{3}{*}{BP02} & 0--1 & 427 & 395 & 1.4\% \\
 & 0--2 & 451 & 420 & 0.4\% \\
 & 1--2 & 346 & 328 & 6.4\% \\
\midrule
\multirow{3}{*}{BP05} & 0--2 & 410 & 387 & 2.1\% \\
 & 0--3 & 430 & 409 & 0.3\% \\
 & 2--3 & 299 & 279 & 11.7\% \\
\bottomrule
\end{tabular}
}
\end{minipage}\hfill
\begin{minipage}[t]{0.48\textwidth}
\centering
\small
\captionof{table}{Per-fold ceiling estimates for LOO across all experiments.}
\label{tab:perfold_ceiling}
\resizebox{\linewidth}{!}{%
\begin{tabular}{llccc}
\toprule
Experiment & Held-out bird & $\sigma{=}50$\,mm & $\sigma{=}100$\,mm & $\sigma{=}150$\,mm \\
\midrule
\multirow{4}{*}{BP01} & Bird 0 & 94.5\% & 79.4\% & 66.4\% \\
 & Bird 1 & 91.0\% & 74.9\% & 62.6\% \\
 & Bird 2 & 94.8\% & 82.3\% & 69.9\% \\
 & Bird 3 & 94.8\% & 82.5\% & 70.3\% \\
\midrule
\multirow{3}{*}{BP02} & Bird 0 & 98.9\% & 91.7\% & 83.0\% \\
 & Bird 1 & 97.4\% & 88.8\% & 78.4\% \\
 & Bird 2 & 98.2\% & 89.8\% & 79.7\% \\
\midrule
\multirow{3}{*}{BP05} & Bird 0 & 94.5\% & 80.8\% & 68.5\% \\
 & Bird 2 & 91.3\% & 78.4\% & 66.9\% \\
 & Bird 3 & 95.2\% & 81.9\% & 69.5\% \\
\bottomrule
\end{tabular}
}
\end{minipage}
\end{table*}

\paragraph{Per-fold ceiling estimates.}
Table~\ref{tab:perfold_ceiling} reports per-fold ceiling estimates for all experiments, computed by adding isotropic Gaussian noise ($\sigma$) to the true vocalizer position and assigning to the nearest candidate bird. BP02 ceilings are substantially higher than BP01 and BP05 due to wider inter-bird separation.

\paragraph{Overlap exclusion.}
Of the 37{,}561 strict valid candidates entering the final geometric release stage, 3{,}784 are removed by overlap filtering. Temporal overlap is therefore the dominant explicit exclusion block within this final stage, and because WhisperSeg proposes candidate windows independently from each bird's accelerometer channel, this overlap exclusion is approximately uniform across birds within each experiment.

\paragraph{Retained vs.\ excluded candidate bias.}
We compared the retention behaviour across birds within each experiment to check whether the curation pipeline is systematically biased. Per-bird retention varies by at most 1--6 percentage points across birds within the same experiment: for BP01 the four birds' retention range is 5.7--6.9\%, for BP02 it is 12.8--18.5\%, and for BP05 it is 26.5--28.3\%. No single bird is disproportionately excluded. The per-experiment retention rate itself varies more substantially (BP01 6.4\%, BP02 16.1\%, BP05 27.6\%), because WhisperSeg over-proposes more aggressively on the longer BP01 recordings (321{,}918 upstream candidates in BP01 vs.\ 90{,}108 in BP02 and 24{,}906 in BP05, from which the 16{,}219 / 11{,}577 / 5{,}829 release sets are ultimately retained); the validation pipeline is what corrects for this.


\section{Baseline Implementation Details}
\label{app:methods}

This appendix provides full architectural, training, and evaluation details for all baselines reported in the main paper.

\subsection{Modality Usage Summary}
\label{app:modality_matrix}

Table~\ref{tab:modality_matrix} summarizes which input modalities each baseline consumes at inference and the attribution strategy it uses.

\noindent
\begin{minipage}{\textwidth}
\centering
\captionof{table}{\textbf{Baseline modality usage and attribution strategy.} \textbf{Audio}: 5-channel microphone array. \textbf{3D}: triangulated bird beak position. \textbf{Head}: head orientation vector. \textbf{Radio}: per-bird FM radio telemetry (summary and/or temporal). \textbf{Video}: synchronized RGB crop tubes. \textbf{Accel}: on-body accelerometer (withheld reference only). A checkmark indicates the modality is consumed at inference; (opt) indicates the model has a variant with and without that input.}
\label{tab:modality_matrix}
\scriptsize
\setlength{\tabcolsep}{3pt}
\begin{tabular*}{\textwidth}{@{\extracolsep{\fill}}p{1.5cm}p{2.15cm}ccccccp{3.0cm}@{}}
\toprule
\textbf{Family} & \textbf{Model} & \textbf{Audio} & \textbf{3D} & \textbf{Head} & \textbf{Radio} & \textbf{Video} & \textbf{Accel} & \textbf{Attribution} \\
\midrule
\multirow{4}{*}{Physics} & LoudestMic & \checkmark & \checkmark & -- & -- & -- & -- & Nearest to loudest mic \\
 & SRP $\pm 20$ & \checkmark & \checkmark & -- & -- & -- & -- & Nearest to SRP peak \\
 & SRP Reranker & \checkmark & \checkmark & \checkmark & \checkmark & -- & -- & Learned reranker \\
 & NeuralSRP & \checkmark & \checkmark & -- & (opt) & -- & -- & Candidate scorer \\
\midrule
Slot-indexed & SELDnet & \checkmark & -- & -- & -- & -- & -- & 4-way slot head \\
\midrule
\multirow{6}{*}{\shortstack[l]{Candidate-\\centric}} & SELDnet++ & \checkmark & \checkmark & \checkmark & \checkmark & -- & -- & Hybrid direct + cand \\
 & SELDnet++ open & \checkmark & \checkmark & \checkmark & \checkmark & -- & -- & Candidate-only matching \\
 & SpatialScorer & \checkmark & \checkmark & -- & (opt) & -- & -- & TDOA $\odot$ audio scorer \\
 & PairRank & \checkmark & \checkmark & \checkmark & \checkmark & -- & -- & Pairwise ranking \\
 & Contrastive & \checkmark & \checkmark & -- & (opt) & -- & -- & Cosine similarity \\
 & Whisper candidate & \checkmark & \checkmark & \checkmark & \checkmark & -- & -- & Candidate scorer \\
\midrule
\multirow{6}{*}{\shortstack[l]{SELD \&\\loc-first}} & ACCDOA & \checkmark & \checkmark & (opt) & (opt) & -- & -- & $-s \cdot \|\hat{p} - p_i\|$ \\
 & EINV2 & \checkmark & \checkmark & (opt) & (opt) & -- & -- & Track--candidate dist \\
 & CST-former & \checkmark & \checkmark & \checkmark & \checkmark & -- & -- & Candidate scorer \\
 & BeamFusion & \checkmark & \checkmark & -- & \checkmark & -- & -- & Beam + transformer \\
 & AudioANN $\times 3$ & \checkmark & \checkmark & -- & -- & -- & -- & Direct / loc / hybrid \\
 & RadioANN $\times 3$ & -- & \checkmark & -- & \checkmark & -- & -- & Direct / loc / hybrid \\
\midrule
\multirow{2}{*}{Vision} & PoseMotion & -- & \checkmark & \checkmark & -- & -- & -- & Per-bird motion score \\
 & VideoCandidate & -- & \checkmark & \checkmark & -- & \checkmark & -- & Per-bird crop score \\
\midrule
Reference & On-body ref. & -- & \checkmark & -- & -- & -- & \checkmark & Per-bird vibration \\
\bottomrule
\end{tabular*}
\end{minipage}

\subsection{Physical Constants and Data Format}
\label{app:constants}

\begin{table}[h]
\centering
\small
\caption{Recording system parameters and physical constants.}
\begin{tabular}{ll}
\toprule
Parameter & Value \\
\midrule
Audio sample rate & 24{,}414.0625~Hz \\
Video frame rate & 47.6837158203125~Hz (= 390625/8192) \\
Samples per video frame & 512 \\
Speed of sound & 343{,}000~mm/s (20$^\circ$C) \\
Cage dimensions (X $\times$ Y $\times$ Z) & 1246 $\times$ 896 $\times$ 1294~mm \\
Cage bounds X (left--right) & $[-90.8, 1155.2]$~mm \\
Cage bounds Y (front--back) & $[-1563.9, -667.9]$~mm \\
Cage bounds Z (bottom--top) & $[236.0, 1530.0]$~mm \\
Coordinate system & Z-up (mm) \\
Number of cage microphones & 5 (Rode NT1, ${\sim}$17~dB gain) \\
Number of nest microphones & 2 (excluded from dataset) \\
Number of radio channels & 4 per experiment \\
Radio tracking range & $\pm$750 / $\pm$700 / $\pm$900~kHz (BP01 / BP02 / BP05) \\
Radio gain & 68~dB \\
Local oscillator frequency & 325 / 350 / 340~MHz (BP01 / BP02 / BP05) \\
\bottomrule
\end{tabular}
\end{table}

\paragraph{Audio clip statistics.}
Clip durations range from 40--784~ms (mean: 150~ms, median: 125~ms) at 24{,}414~Hz. Each clip contains the full vocalization from onset to offset plus $\pm$10~ms padding.

\paragraph{Video frame statistics.}
Each sample includes 4--38 video frames (mean: 9.1) from the composite multi-view video (2976$\times$2016 pixels, 47.68~fps).

\paragraph{Position normalization.}
All 3D positions are normalized to $[0, 1]$ by: $\mathbf{p}_\text{norm} = (\mathbf{p} - \mathbf{p}_\text{lo}) / (\mathbf{p}_\text{hi} - \mathbf{p}_\text{lo})$, where $\mathbf{p}_\text{lo}$ and $\mathbf{p}_\text{hi}$ are the cage bounds listed above. Denormalization reverses this mapping.

\paragraph{Bird masking.}
BP01 and BP05 have 4 birds; BP02 has 3 birds. The data loader pads arrays to the maximum bird count in each batch, with a boolean mask indicating valid birds. Padded positions, head orientations, and radio features are set to zero; logits at padded positions are set to $-\infty$ before softmax.

\paragraph{Radio feature summary.}
The 12-dimensional per-bird radio summary consists of:
\begin{itemize}[nosep]
    \item Mean and standard deviation (over frames in the window) of the baseband signal power (dBFS) at the four antennas and the summed master vector: \texttt{powRA}, \texttt{powRB}, \texttt{powRC}, \texttt{powRD}, \texttt{powRM} (10 features).
    \item Standard deviation of the PLL tracking frequency \texttt{frq} (1 feature).
    \item Standard deviation of the master-vector phase angle \texttt{thetaS} ($\theta_i = \arg u_M$, in degrees; 1 feature).
\end{itemize}
All 21 raw radio signals per bird per frame are available in the parquet files for custom feature extraction. Using BirdRadio's naming~\citep{ruttimann2024birdpark}, the 21 signals are: the PLL tracking frequency \texttt{frq}; the master-vector phase \texttt{thetaS} ($\theta_i$) and the per-antenna rotated-vector angles \texttt{thetaA}--\texttt{thetaD} ($\alpha_i^a$); the per-antenna compensation phases \texttt{phiA}--\texttt{phiD} ($\phi_i^a$) and their sum \texttt{phiM}; the baseband signal powers \texttt{powRA}--\texttt{powRD},\texttt{powRM} and variances \texttt{powNA}--\texttt{powND},\texttt{powNM} (all in dBFS, at the sender frequency). None of these is an angle of arrival.

\subsection{Classical Baselines (No Training)}

\paragraph{Random baseline.}
For each sample, the predicted vocalizer is drawn uniformly at random from the $N$ candidate birds present in the scene. Expected accuracy: $1/N$ (${\sim}25\%$ for 4 birds, ${\sim}33\%$ for 3 birds).

\paragraph{LoudestMic.}
LoudestMic is a deterministic energy-plus-geometry heuristic. For each event, per-channel root-mean-square energy is computed over the audio window and the loudest microphone is identified:
\begin{equation}
c^\star = \arg\max_{c \in \{0,\ldots,4\}} \sqrt{\tfrac{1}{T} \sum_{t} x_c(t)^2},
\end{equation}
where $x_c(t)$ is the waveform on mic $c$. The predicted vocalizer is the visible bird whose 3D position $\mathbf{p}_b$ is closest to the loudest microphone's 3D position $\mathbf{m}_{c^\star}$:
\begin{equation}
\hat{b} = \arg\min_b \lVert \mathbf{p}_b - \mathbf{m}_{c^\star} \rVert.
\end{equation}
No time-difference information, voice identity, or training is used. This baseline sits between Random and SRP in the deterministic-spatial family: it exploits which mic heard the call loudest (a proxy for proximity) but, unlike SRP, it cannot resolve configurations where multiple birds are similarly close to the same microphone.

\paragraph{SRP $\pm$20.}
Steered Response Power with Phase Transform (SRP-PHAT)~\citep{dibiase2000srp} is our strongest classical baseline. For each of the $\binom{5}{2} = 10$ microphone pairs, we compute the GCC-PHAT cross-correlation~\citep{knapp1976gccphat}:
\begin{equation}
R_{ij}(\tau) = \text{IFFT}\left(\frac{X_i(f) \cdot X_j^*(f)}{|X_i(f) \cdot X_j^*(f)| + \epsilon}\right)
\end{equation}
where $X_i(f)$ is the FFT of the $i$-th microphone channel and $\epsilon = 10^{-10}$.

For each candidate bird $b$ at 3D position $\mathbf{p}_b$, we compute the expected TDOA for each microphone pair $(i,j)$:
\begin{equation}
\tau_{ij}^{(b)} = \frac{\|\mathbf{p}_b - \mathbf{m}_i\| - \|\mathbf{p}_b - \mathbf{m}_j\|}{c}
\end{equation}
where $\mathbf{m}_i$ is the $i$-th microphone position and $c = 343{,}000$~mm/s is the speed of sound.

The SRP score for bird $b$ is:
\begin{equation}
S(b) = \sum_{(i,j)} \max_{\delta \in [-W, W]} R_{ij}\!\left(\tau_{ij}^{(b)} + \delta\right)
\end{equation}
where $W = 20$ samples (${\sim}820$~$\mu$s) is the search window that compensates for reverberant TDOA bias. The predicted vocalizer is $\arg\max_b S(b)$.

SRP requires no training. It uses the known 3D microphone positions (Section~3.1) and the candidate bird positions from the vision pipeline. The $\pm 20$ search window was selected based on analysis of the GCC-PHAT peak distribution (see main paper ceiling estimate).

Audio preprocessing: each channel is zero-mean and unit-variance normalized before FFT.

\subsection{SELDnet (Audio-Only Learned Baseline)}
\label{app:seldnet}

SELDnet is a convolutional-recurrent audio pipeline adapted from the original SELDnet formulation~\citep{adavanne2018seldnet} and later DCASE SELD baselines~\citep{politis2022dcase} to event-local candidate attribution.

\paragraph{Label space.}
SELDnet is a slot-indexed attribution model. Its output head predicts one of up to four event-local candidate slots, corresponding to the \texttt{vocalizer\_idx} $\in \{0,1,2,3\}$ field in the release. This index is not a global bird identity. Within a given experiment the slot-to-bird mapping is stable across events, so a slot classifier can effectively memorize the vocal signatures of that experiment's birds. Across experiments the same slot index refers to different physical birds, and in within-experiment leave-one-bird-out evaluation the held-out bird's slot is absent from training entirely. A pure slot classifier therefore fails by construction under LOO, which explains the zero-accuracy behavior of SELDnet and other slot-indexed baselines.

\paragraph{Architecture.}
\begin{itemize}[nosep]
    \item \textbf{Input:} 5-channel mel spectrogram ($N_\text{mics} \times N_\text{mels} \times T$).
    \item \textbf{Mel extraction:} 80-bin log-mel spectrogram at 24{,}414~Hz, $n_\text{fft} = 1024$, hop length 256 samples, $f_\text{min} = 500$~Hz, $f_\text{max} = 12{,}000$~Hz.
    \item \textbf{CNN:} Three convolutional blocks, each consisting of Conv2d $\to$ BatchNorm2d $\to$ ReLU $\to$ MaxPool2d(4,1). Channel progression: $5 \to 64 \to 128 \to 256$. Kernel size: $3 \times 3$, padding 1. Pooling reduces the frequency dimension by $4\times$ per block ($80 \to 20 \to 5 \to 1$).
    \item \textbf{GRU:} 2-layer bidirectional GRU with hidden size 128, yielding 256-dim output per time step. Dropout 0.1 between layers.
    \item \textbf{Pooling:} Mean pooling over the time dimension.
    \item \textbf{Slot-indexed attribution head:} Linear(256, 128) $\to$ ReLU $\to$ Dropout(0.2) $\to$ Linear(128, $N_\text{max\_slots}$), where $N_\text{max\_slots} = 4$ is the maximum number of visible candidates in any event. At inference a per-event candidate mask zeroes out slot logits for unused positions.
    \item \textbf{DOA head:} Linear(256, 64) $\to$ ReLU $\to$ Linear(64, 3) $\to$ Sigmoid (used as an auxiliary source-position head).
\end{itemize}

\paragraph{Parameters.} ${\sim}$1.03M total.

\paragraph{Audio preprocessing.}
Each channel is independently zero-mean, unit-variance normalized. Mel spectrograms are instance-normalized: per-sample, per-channel zero-mean and unit-variance over frequency and time dimensions.

\subsection{PairRank and SpatialScorer}
\label{app:pairrank_spatial}

\paragraph{PairRank.}
\textbf{PairRank} removes the fixed class head and instead learns pairwise candidate preferences. The model first computes GCC-PHAT summary statistics for each of the 10 microphone pairs (peak, mean, and sharpness), producing a 30-dimensional audio summary. This summary is mapped through an MLP to a 96-dimensional audio embedding. Each candidate bird is encoded from normalized 3D position (3), head orientation (3), radio summary (12), and expected TDOA values for the 10 microphone pairs, again using a 96-dimensional MLP. For every candidate pair $(i,j)$, the model forms a pair feature containing the two candidate embeddings, their relative displacement, their Euclidean distance, and the mean difference in expected TDOA. A pairwise scorer predicts whether bird $i$ is more plausible than bird $j$, and the aggregated pairwise wins become the final candidate scores. Localization is candidate-conditioned: the predicted position is the softmax-weighted average of candidate positions.

\paragraph{SELDnet++.}
\textbf{SELDnet++} extends SELDnet with full multimodal candidate conditioning. The audio encoder is identical to SELDnet (5-channel mel-spectrogram $\to$ 3-layer CNN with frequency pooling $\to$ 2-layer BiGRU $\to$ temporal mean pool $\to$ 256-dim embedding). Each candidate bird is encoded from the concatenation of its 3D position (3), head orientation (3), radio summary (12), and a radio temporal encoding (7-channel Conv1d $\to$ BN $\to$ ReLU $\to$ Conv1d $\to$ BN $\to$ ReLU $\to$ AdaptiveAvgPool1d $\to$ 64), yielding an 82-dimensional input that is mapped through a 2-layer MLP to a 128-dimensional candidate embedding. The audio embedding is projected to 128 dimensions, and per-candidate scores are computed as $10 \cdot \langle \hat{\mathbf{a}}, \hat{\mathbf{c}}_i \rangle$ (normalized dot product). A slot-indexed classification head (Linear $\to$ ReLU $\to$ Dropout $\to$ Linear) with the same 4-way output as SELDnet is blended with the candidate scores via a learnable sigmoid gate $\beta$, initialized at 0.5; \textbf{SELDnet++ open} zeros this gate and keeps only the candidate-scoring branch, forcing attribution to match the audio embedding against candidate state without any slot-indexed shortcut. SELDnet++ is the only model in the suite that combines the SELDnet audio backbone with all four candidate-side modalities (position, orientation, radio summary, and radio temporal).

\paragraph{SpatialScorer.}
\textbf{SpatialScorer} is the strongest purely spatial candidate model in the current suite. It uses GCC-PHAT lag features only, intentionally excluding mel spectrograms and timbre-heavy audio summaries. The 10 microphone-pair cross-correlations are cropped to a $\pm 60$ sample lag window and stacked into a $(10 \times 121 \times T)$ tensor. A CNN encoder maps this GCC volume to a 256-dimensional audio embedding. In parallel, each candidate bird is encoded from its expected TDOA vector across the same 10 microphone pairs, and optional radio context is obtained from a temporal radio encoder. Candidate scores are produced from the concatenation of the audio embedding, candidate TDOA embedding, their elementwise product, and optional radio features. We additionally use a gradient-reversal adversarial head~\citep{ganin2016domain} to discourage experiment-specific shortcuts, making this baseline explicitly transfer-oriented rather than slot-indexed.

\subsection{Localization-First AudioANN and RadioANN Baselines}
\label{app:audio_radio_ann}

\paragraph{AudioANN family.}
\textbf{AudioANN} uses a shared audio backbone built from a 64-bin multichannel mel front-end, a 3-layer CNN ($5 \to 64 \to 128 \to 128$ channels), and a 2-layer bidirectional LSTM with hidden size 128. The pooled audio representation is used in three ways:
\begin{itemize}[nosep]
    \item \textbf{AudioANN-direct:} a direct MLP predicts candidate logits.
    \item \textbf{AudioANN-locfirst:} a position head predicts 3D source position, and candidate logits are the negative distances between that position and each candidate bird.
    \item \textbf{AudioANN-hybrid:} a learnable interpolation blends direct logits and localization-derived logits.
\end{itemize}

\paragraph{RadioANN family.}
\textbf{RadioANN} is the radio-only analogue. Its backbone is a Conv1d encoder over the 7 raw radio temporal channels per bird, with channel progression $7 \to 64 \to 128$ followed by global temporal pooling. Again we evaluate three variants:
\begin{itemize}[nosep]
    \item \textbf{RadioANN-direct:} score each bird directly from its radio embedding.
    \item \textbf{RadioANN-locfirst:} predict a per-bird 3D position from the radio embedding and rank candidates by distance.
    \item \textbf{RadioANN-hybrid:} blend direct radio scores with localization-derived scores.
\end{itemize}
These models are intentionally simple and expose how much attribution signal is available from straightforward audio or radio encoders before adding stronger spatial inductive bias.

\subsection{Vision-Only Candidate Baselines}
\label{app:vision}

\paragraph{PoseMotion.}
\textbf{PoseMotion} is the pose-only control. For each bird and frame we derive a compact motion vector from the tracked 3D keypoints: relative keypoint coordinates together with frame-to-frame deltas. A 2-layer MLP projects each frame to a hidden representation, and a 2-layer bidirectional GRU pools the short temporal window into a per-candidate motion embedding. Candidate logits are then predicted from the motion embedding concatenated with the bird's normalized 3D position and head orientation. Localization is candidate-conditioned: the predicted source position is the softmax-weighted average of candidate positions. This baseline measures whether short-window beak/head/backpack motion alone is sufficient for caller attribution.

\paragraph{VideoCandidate.}
\textbf{VideoCandidate} is the RGB active-speaker-style control. We extract synchronized top-view and back-view crop tubes around each visible bird from the composite video, using a $32\times32$ crop budget and 4 sampled frames per event window. Each frame is encoded by a small CNN ($3 \to 32 \to 64 \to 96$ channels with strided convolutions), the two views are fused by an MLP, and a 2-layer bidirectional GRU pools the crop tube over time. Candidate logits are again produced from the fused visual embedding together with the bird's normalized 3D position and head orientation. Localization is candidate-conditioned via the score-weighted candidate position. This row is the closest benchmark analogue to classical active-speaker baselines such as AVA-style, TalkNet-style, and UniTalk-style visual speaker attribution~\citep{roth2020ava,tao2021someone,nguyen2025unitalk}, but adapted to multi-bird candidate scoring rather than person-centric diarization.

\subsection{Literature-Style SELD Adaptations}
\label{app:lit_seld}

\paragraph{Shared spatial encoder.}
The \textbf{ACCDOA}, \textbf{EINV2}, and \textbf{CST-former} rows all begin from SALSA-Lite spatial audio features~\citep{nguyen2022salsalite} and normalized candidate state. The convolutional variants use a shared spatial encoder composed of SALSA-Lite extraction followed by a CNN with channel progression $5 \to 64 \to 128 \to 192$; the transformer variant uses a 3-layer SpatialConformer in the spirit of Conformer-style encoder blocks~\citep{gulati2020conformer} with 4 attention heads and model width 192.

\paragraph{ACCDOA adaptation.}
Our \textbf{ACCDOA} adaptation follows the Multi-ACCDOA formulation~\citep{shimada2022multiaccdoa}: it applies a 2-layer bidirectional GRU on top of the spatial encoder, predicts a framewise activity scalar, and regresses a framewise 3D position vector. The final source position is the activity-weighted average over frames. Attribution is then performed by converting distances between the predicted position and candidate bird positions into logits, optionally with an additional radio-conditioned position-bias scorer.

\paragraph{EINV2 adaptation.}
\textbf{EINV2} retains the event-independent trackwise formulation~\citep{cao2021einv2}. A recurrent spatial encoder produces a pooled representation that predicts $K=3$ latent tracks, each with an activity logit and 3D position. Candidate attribution is obtained by matching candidate positions to the latent tracks through distance-based logits and taking the best track per candidate. This produces a free-localization model rather than a pure candidate-conditioned scorer.

\paragraph{CST-former adaptation.}
\textbf{CST-former} uses a transformer-style spatial encoder over SALSA-Lite features~\citep{shul2024cstformer}. The pooled transformer representation drives both a 3D position head and a candidate scorer that consumes the audio embedding together with each bird's position, head orientation, and optional radio context. It therefore sits between a free localization model and a candidate-conditioned scorer.

\subsection{Physics-Guided Learned Spatial Baselines}
\label{app:physics_learned}

\paragraph{SRP reranker.}
\textbf{SRP reranker} is the simplest learned extension of SRP. We first precompute frozen SRP-style features for every candidate: for each microphone pair we record the GCC-PHAT peak within a $\pm20$ sample window around the expected TDOA, the narrow-window mean, and a peak-to-mean sharpness statistic. Concatenating these over 10 microphone pairs yields a 30-dimensional SRP descriptor per candidate. A small 3-layer MLP then scores each candidate from this descriptor plus candidate position, head orientation, and the 12-dimensional radio summary. The model therefore tests whether a small learned reranker can improve on physics without replacing it.

\paragraph{NeuralSRP.}
\textbf{NeuralSRP} keeps the same geometry-first philosophy but learns the spatial representation end to end. The input is a SALSA-Lite tensor composed of per-microphone log spectrograms and normalized inter-channel phase differences~\citep{nguyen2022salsalite} (9 channels total). A ResNet-style encoder with squeeze-excitation blocks~\citep{he2016deep,hu2018senet} and frequency-aware normalization maps these features to a 256-dimensional audio embedding. Each candidate bird is represented by its expected TDOA vector relative to microphone 0, encoded by an MLP; optional radio temporal features are encoded by a small Conv1d stack. Attribution uses a learned per-candidate scorer over audio, TDOA, and optional radio embeddings, while localization comes from a separate MLP head on the shared audio embedding.

\paragraph{BeamFusion.}
\textbf{BeamFusion} adds explicit source-focused filtering. For each candidate bird, a learned beamformer computes physics steering vectors from the candidate 3D position and microphone geometry, refines them through an MLP, and applies them to the multichannel STFT to produce a beamformed per-bird spectrogram. These per-bird queries are fused with a global SALSA-Lite + SpatialConformer context stream through a transformer decoder, again using a Conformer-style spatial encoder~\citep{gulati2020conformer}. Optional radio summaries are concatenated to the candidate queries before decoding. This is the most explicit multichannel ``focus on each candidate position first, then classify'' baseline in the suite.

\paragraph{Contrastive candidate model.}
\textbf{Contrastive} learns a shared embedding space for audio clips and candidate spatial signatures. The audio encoder is a multichannel mel CNN followed by a 2-layer bidirectional GRU and a projection to a 128-dimensional L2-normalized embedding. Each candidate is encoded from TDOA features across all microphone pairs, inverse-distance energy patterns to the microphones, normalized 3D position, and optional temporal radio embeddings. Attribution uses cosine similarity divided by a temperature parameter. This baseline is explicitly designed for transfer: it never learns a fixed slot-indexed class mapping.

\paragraph{Whisper candidate scorer.}
\textbf{Whisper candidate} uses a frozen Whisper encoder~\citep{radford2023whisper} as a pretrained audio backbone. Each microphone channel is resampled to 16~kHz, encoded independently by Whisper, pooled with a VocSim-style mean-over-time and mean-over-feature operation~\citep{basha2024vocsim}, and fused across microphones to a 256-dimensional audio embedding. Candidate features contain bird position, head orientation, and optional radio summary; a small MLP combines the audio embedding, candidate embedding, and their elementwise interaction to score candidates. This baseline measures whether a stronger pretrained audio representation materially changes transfer.

\subsection{Withheld On-Body Reference}
\label{app:oracle}

\paragraph{Withheld on-body reference.}
The \textbf{withheld on-body reference} uses only the per-bird accelerometer waveform provided by the validated on-body sensor. Each bird's accelerometer trace is converted to a log-power spectrogram with $n_\text{fft}=256$ and hop length 64. A shared 2D CNN with channel progression $1 \to 32 \to 64 \to 96 \to 96$ and max-pooling produces a per-bird feature map. Mean and max pooled features are concatenated and mapped through a small MLP to one score per bird. The predicted 3D position is the position of the predicted bird. This model is reported only as a ceiling for candidate-conditioned attribution and localization, not as a benchmark competitor.

\subsection{Training Protocol (All Learned Models)}
\label{app:training}

Unless otherwise stated, all learned baselines are trained with AdamW and a cosine-annealing schedule. Model-family-specific settings are summarized in Table~\ref{tab:shared_hyperparams}; this section lists the conventions shared across families.

\paragraph{Optimizer and schedule.}
All learned baselines use AdamW with cosine annealing. SELDnet uses learning rate $2 \times 10^{-4}$ and weight decay $5 \times 10^{-3}$. PairRank, SpatialScorer, the literature-style SELD baselines, AudioANN/RadioANN, BeamFusion, Contrastive, and Whisper-candidate use learning rate $2 \times 10^{-4}$ and weight decay in the range $10^{-3}$ to $10^{-2}$. SRP reranker uses a larger learning rate of $10^{-3}$ because it only trains a small MLP on frozen SRP descriptors. SELDnet is trained for 60 epochs on the session split and 40 epochs on LOO-style controls. Most matrix baselines are trained for shorter budgets because they are swept over many more cells: PairRank, SpatialScorer, ACCDOA, EINV2, CST-former, AudioANN/RadioANN, BeamFusion, and Contrastive use 8 epochs; Whisper-candidate uses 6 epochs; NeuralSRP uses 8 epochs; SRP reranker uses 40 epochs.

\paragraph{Loss.}
$\mathcal{L} = \mathcal{L}_\text{CE} + 0.1 \cdot \mathcal{L}_{L_2}$ for models with an explicit localization head, where:
\begin{itemize}[nosep]
    \item $\mathcal{L}_\text{CE}$: Cross-entropy with label smoothing $\epsilon = 0.1$.
    \item $\mathcal{L}_{L_2}$: Mean squared error between predicted and true vocalizer position (normalized coordinates).
\end{itemize}

\paragraph{Data augmentation.} Applied with the following probabilities per batch:
\begin{itemize}[nosep]
    \item \textbf{Channel dropout} (25\%): 1--2 randomly selected microphone channels are scaled by $U(0, 0.1)$.
    \item \textbf{Per-channel gain variation} (30\%): Each channel is multiplied by an independent gain factor $U(0.5, 2.0)$.
    \item \textbf{Additive noise} (40\%): Gaussian noise with standard deviation $U(0.002, 0.02)$ is added to all channels.
\end{itemize}

\paragraph{Gradient clipping.} Maximum gradient norm of 1.0.

\paragraph{Batch size.}
Batch size depends on model cost. SELDnet uses 64. PairRank and SRP reranker use 32--128. SpatialScorer and the literature-style SELD baselines use 16. AudioANN/RadioANN use 32. Whisper-candidate uses 8 because the frozen Whisper encoder dominates memory.

\paragraph{Seeds.}
The matrix-style transfer families (PairRank, SpatialScorer, ACCDOA, EINV2, CST-former, AudioANN/RadioANN, BeamFusion, Contrastive, and Whisper-candidate), together with SELDnet, NeuralSRP, and the withheld on-body reference, are reported as mean $\pm$ standard deviation over 3 seeds $\{42,123,456\}$. The SRP reranker uses 5 seeds $\{42, 123, 456, 789, 1337\}$; SRP~$\pm20$, LoudestMic, and Random are deterministic and reported without seed variation.

\paragraph{Early stopping.}
Early stopping is not used. All models are trained for the full epoch budget and the best validation-accuracy checkpoint is selected.

\paragraph{Compute resources.}
All reported training runs were executed on a single local workstation with one NVIDIA RTX~3090 GPU (24\,GB VRAM); most scripts used up to 4 CPU data-loader workers. Summing successful stage durations from the saved logs, the reported main benchmark matrix required approximately 42.0 GPU-hours end-to-end. The completed comparable diagnostics add 6.6 hours, and the two vision-only controls add 1.3 hours (PoseMotion) and 3.3 hours (VideoCandidate, including offline crop-cache precomputation), for a total of about 53.1 GPU-hours across the experiments reported in the paper. On a single 3090 this corresponds to roughly 2.2 wall-clock days for the retained experiments alone.

\paragraph{Runtime variation across families.}
Most retained matrix families fall in the 1--2 hour range when their three benchmark regimes are run end-to-end (e.g., PairRank 1.1\,h, SpatialScorer 1.3\,h, ACCDOA 1.1\,h, EINV2 1.4\,h, CST-former 1.6\,h, Contrastive 1.1\,h, Whisper candidate 2.3\,h). The main outliers are the heavier LOO-sensitive or preprocessing-heavy families: benchmark-family ablations (5.4\,h), AudioANN/RadioANN (5.4\,h), SRP reranker (7.9\,h), NeuralSRP (3.4\,h), and BeamFusion (8.9\,h). Across successful matrix stages, the median stage duration is 35.4 minutes, with a range from 14.9 minutes to 4.1 hours.

\begin{figure}[h]
\centering
\includegraphics[width=\textwidth]{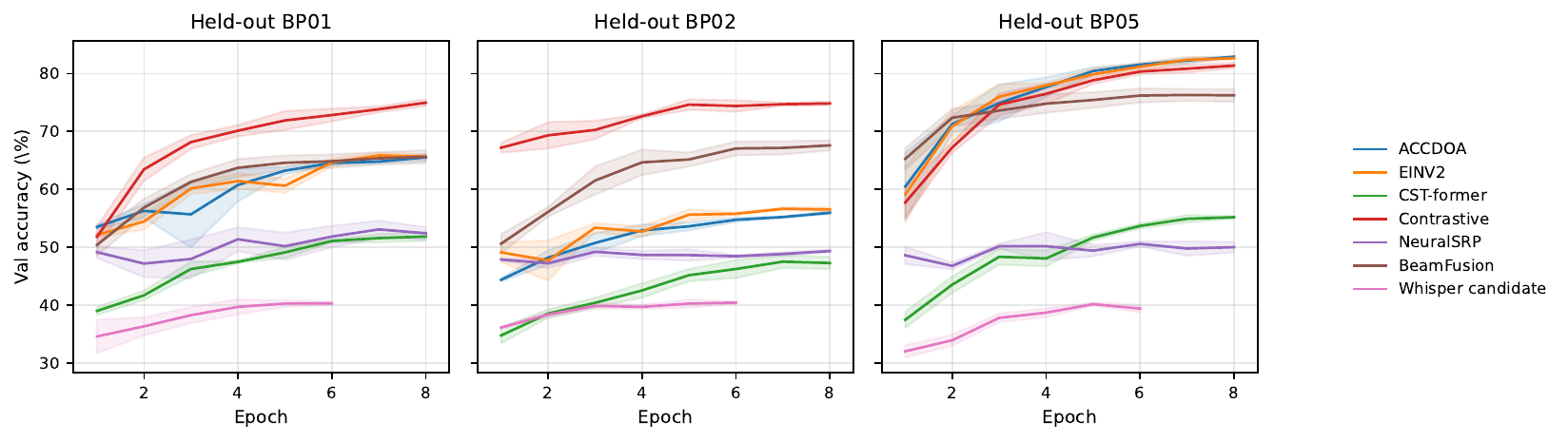}
\caption{\textbf{Validation accuracy per epoch for matrix-family baselines on the three held-out-experiment splits.} Lines are seed means, shaded bands are $\pm 1$ standard deviation across the three training seeds.}
\label{fig:learning_curves}
\end{figure}

\paragraph{Learning curves.}
Figure~\ref{fig:learning_curves} shows validation accuracy as a function of training epoch for the seven matrix-family baselines whose training logs contain per-epoch validation metrics, under the three held-out-experiment splits. Shaded bands are one standard deviation across the three training seeds. Most models are close to their final validation accuracy by the end of the 8-epoch budget used in the main sweep, but CST-former has not reached a plateau by epoch 8 on held-out BP05; Appendix~\ref{app:budget} tests the effect of the budget directly. SpatialScorer, PairRank, and the withheld on-body reference log only best-epoch summaries and are omitted from the per-epoch plot.

\subsection{Training Budget Control}
\label{app:budget}
To test whether the 8-epoch budget understated CST-former and Whisper-candidate, we reran both for 60 epochs with the same architecture, preprocessing, optimizer, splits, seeds, and validation-only checkpoint selection, and ran ACCDOA under the same extension as a control (Table~\ref{tab:budget}). CST-former's selected epochs span 23--55 (median 44), and its LOO accuracy rises from 15.0\% to $19.9\pm5.7$\%, so its submitted result is budget-sensitive. Whisper-candidate does not improve, and ACCDOA changes by 1.3 points LOO and 1.5 points session-disjoint. Training budget therefore affected CST-former's absolute score but did not create the formulation-level pattern.

\begin{table}[h]
\centering
\small
\caption{\textbf{Attribution accuracy (\%) with the submitted budget and with 60 epochs.} Held-out exp.\ mean averages BP01, BP02, and BP05.}
\label{tab:budget}
\begin{tabular}{llrr}
\toprule
Method & Evaluation aggregate & Submitted & 60-epoch \\
\midrule
CST-former & LOO & 15.0 & $19.9 \pm 5.7$ \\
CST-former & Held-out exp.\ mean & 30.0 & 33.7 \\
Whisper-candidate & LOO & 12.1 & $11.7 \pm 1.9$ \\
Whisper-candidate & Held-out exp.\ mean & 31.4 & 30.3 \\
ACCDOA & LOO & 55.4 & 56.7 \\
ACCDOA & Session-disjoint & 65.6 & 67.1 \\
\bottomrule
\end{tabular}
\end{table}

\paragraph{Configuration choices.}
Configurations were fixed once and reused across all folds and seeds, with architectures following the cited formulations and optimizer settings taken from the shared defaults of Table~\ref{tab:shared_hyperparams}; no setting was tuned per method. ACCDOA, EINV2, and CST-former keep their published output formulations on a shared spatial encoder (192-dimensional, 64 mel bands) rather than their published model sizes. Checkpoints are selected on validation accuracy only; no test result selected a configuration, budget, or checkpoint. Table~\ref{tab:shared_hyperparams} lists each family's optimizer settings, budget, batch size, seeds, and label smoothing, and the loss weights are given below it.

\subsection{Evaluation Protocol}
\label{app:eval}

\paragraph{Session-disjoint split.}
Train: 19{,}365 samples. Val: 6{,}869 samples. Test: 7{,}391 samples. By experiment, the session split allocates BP01 as 12{,}881/1{,}008/2{,}330, BP02 as 4{,}455/4{,}155/2{,}967, and BP05 as 2{,}029/1{,}706/2{,}094 for train/val/test.

\paragraph{Held-out experiment.}
We rotate the held-out experiment across BP01, BP02, and BP05. In each fold, training uses the remaining two experiments and testing uses the held-out one. The resulting train/val/test sizes are 6{,}484/5{,}861/16{,}219 for held-out BP01, 14{,}910/2{,}714/11{,}577 for held-out BP02, and 17{,}336/5{,}163/5{,}829 for held-out BP05. This is the main transfer regime in the paper because it combines identity shift with experiment-level nuisance changes such as year, cohort, and recording date.

\paragraph{Leave-one-bird-out (LOO).}
We run within-experiment LOO controls for all experiments. BP01 has 4 folds with train sizes 9{,}418--12{,}535, validation sizes 1{,}066--1{,}407, and test sizes 2{,}277--5{,}735. BP02 has 3 folds with train sizes 5{,}989--8{,}569, validation sizes 665--943, and test sizes 2{,}065--4{,}923. BP05 has 3 folds with train sizes 1{,}974--5{,}201, validation sizes 222--572, and test sizes 56--3{,}633. Same recording dates appear in both train and test; only the bird identity changes.

\paragraph{Metrics.}
\begin{itemize}[nosep]
    \item \textbf{Top-1 accuracy:} Fraction of correctly attributed vocalizations.
    \item \textbf{Macro-F1:} Unweighted mean of per-class F1 scores.
    \item \textbf{Localization error:} Euclidean error in 3D source position, reported separately for free localization and candidate-conditioned localization.
\end{itemize}

\paragraph{Shuffle intervention.}
For the modality-shuffle diagnostic, we run 10 independent random shuffles (permuting bird indices within each sample for the specified modality) and report mean $\pm$ std of test accuracy. Audio is never shuffled. Shuffling destroys the correspondence between a bird's position/radio features and its identity while preserving the marginal distribution of feature values.

\subsection{Training Hyperparameters}
\label{app:hyperparams}

Table~\ref{tab:shared_hyperparams} summarizes training hyperparameters across all learned baselines.

\begin{table}[h]
\centering
\small
\caption{Shared hyperparameters across learned baselines. \textbf{Architecture}: source of the model design (ours = introduced in this work). Optimizer settings are shared defaults and were not tuned per method.}
\label{tab:shared_hyperparams}
\resizebox{\textwidth}{!}{%
\begin{tabular}{llcccccc}
\toprule
Family & Architecture & LR & WD & Epochs & Batch & Seeds & Label smooth \\
\midrule
SELDnet & CRNN of \citet{adavanne2018seldnet} & $2{\times}10^{-4}$ & $5{\times}10^{-3}$ & 60 & 64 & 3 & 0.1 \\
SRP reranker & ours & $1{\times}10^{-3}$ & $1{\times}10^{-2}$ & 40 & 128 & 5 & 0.05 \\
NeuralSRP & SALSA-Lite input \citep{nguyen2022salsalite} & $2{\times}10^{-4}$ & $5{\times}10^{-3}$ & 8 & 32 & 3 & 0.1 \\
PairRank & ours & $2{\times}10^{-4}$ & $5{\times}10^{-3}$ & 8 & 32 & 3 & 0.03 \\
SpatialScorer & ours & $2{\times}10^{-4}$ & $5{\times}10^{-3}$ & 8 & 16 & 3 & 0.1 \\
Contrastive & ours & $2{\times}10^{-4}$ & $5{\times}10^{-3}$ & 8 & 16 & 3 & 0.1 \\
Whisper candidate & frozen Whisper \citep{radford2023whisper} & $2{\times}10^{-4}$ & $1{\times}10^{-3}$ & 6 & 8 & 3 & 0.0 \\
ACCDOA / EINV2 / CST-former & cited output heads, shared encoder & $2{\times}10^{-4}$ & $5{\times}10^{-3}$ & 8 & 16 & 3 & 0.0 \\
BeamFusion & ours & $2{\times}10^{-4}$ & $5{\times}10^{-3}$ & 8 & 16 & 3 & 0.1 \\
AudioANN / RadioANN & ours & $2{\times}10^{-4}$ & $5{\times}10^{-3}$ & 8 & 32 & 3 & 0.1 \\
PoseMotion & ours & $2{\times}10^{-4}$ & $5{\times}10^{-3}$ & 8 & 32 & 3 & 0.1 \\
VideoCandidate & ours & $2{\times}10^{-4}$ & $5{\times}10^{-3}$ & 8 & 16 & 3 & 0.1 \\
On-body ref. & ours & $2{\times}10^{-4}$ & $5{\times}10^{-3}$ & 6 & 32 & 3 & 0.0 \\
\bottomrule
\end{tabular}}
\end{table}

All learned models use AdamW with cosine-annealing schedule and gradient clipping at max norm 1.0. Loss is $\mathcal{L}_\text{CE} + \lambda_\text{loc} \cdot \mathcal{L}_{L_2}$ unless otherwise noted, where $\lambda_\text{loc}$ varies by model (default 0.1; AudioANN/RadioANN use 0.5; Whisper candidate uses 0.05; SELD literature baselines use 0.05; SRP reranker and withheld on-body reference use pure cross-entropy).

\subsubsection{Shared Sub-Modules}
\label{app:arch_shared}

\paragraph{JointRadioTemporalEncoder.}
Used by ACCDOA (radio bias scorer), EINV2, and CST-former radio-context variants.
\begin{itemize}[nosep]
    \item \textbf{Input:} $(B, N, 7, T_\text{frames})$ raw radio signals for all birds.
    \item \textbf{Joint CNN:} Three Conv2d layers over the bird$\times$time grid: Conv2d$(7 \to 64, (3,5))$ $\to$ BN $\to$ ReLU $\to$ Conv2d$(64 \to 64, (3,3))$ $\to$ BN $\to$ ReLU $\to$ Conv2d$(64 \to 64, (3,3))$ $\to$ BN $\to$ ReLU. Mean-pool over time $\to$ per-bird 64-dim.
    \item \textbf{Cross-bird attention:} 1-layer Transformer encoder (4 heads, FFN dim 128, pre-norm). Learnable bird-position embeddings (up to 8 birds).
    \item \textbf{Outputs:} Per-bird embedding: $[\mathbf{b}_i; \bar{\mathbf{b}}]$ concatenated with global context $\to$ Linear(128, 64) $\to$ ReLU $\to$ Linear(64, 64). Global embedding: Linear(64, 64) $\to$ ReLU $\to$ Linear(64, 64).
\end{itemize}

\paragraph{SALSALiteExtractor.}
Shared front-end for all SELD-literature baselines, NeuralSRP, and BeamFusion context stream.
\begin{itemize}[nosep]
    \item Channel 0: log-mel spectrogram of the reference mic.
    \item Channels 1--$(M{-}1)$: normalized inter-channel phase differences (NIPDs) relative to mic~0.
    \item Output: $(B, M, T, F)$ where $M = n_\text{mics}$ and $F = n_\text{mels}$.
\end{itemize}


\section{Descriptive and Comparative Statistics}
\label{sec:appendix_stats}

This appendix reports descriptive and comparative statistics that support the interpretation of the benchmark results: detailed attribution analysis, intervention diagnostics, fold-level leave-one-bird-out performance, sample-weighted versus unweighted LOO summaries, localization results, and global transfer rankings across tasks.

\subsection{Attribution: Detailed Analysis}
\label{sec:appendix_attribution_detail}

This subsection expands the main-text \textsc{Attribution} paragraph (§\ref{sec:experiments}).

\paragraph{Slot-collapse failure.}
SELDnet++ achieves the highest session score (86.5\%) yet collapses to 0\% LOO. The failure is diagnostic rather than accidental: although SELDnet++ includes a candidate-matching branch, training is dominated by its slot-indexed head, which can exploit stable within-experiment slot-to-bird mappings but cannot generalize when the held-out bird's slot is absent during training. Removing that head in SELDnet++ open restores non-zero LOO (9.3\%) at a large session cost (61.5\%), confirming that the shortcut is both attractive and brittle; SELDnet (81.8\%) and AudioANN-direct (72.1\%) exhibit the same slot-collapse failure. SELDnet++ open is informative on its own: forced onto pure candidate matching with full candidate conditioning (3D position, head direction, radio), the SELDnet backbone lands at 61.5\% session / 9.3\% LOO, well below the geometry-first rows despite having access to the same candidate-side modalities, indicating that the architecture does not on its own convert candidate features into transferable attribution.

\paragraph{Regime-level results.}
Across regimes the common ingredient of the strongest rows is not a single backbone but a formulation: the highest observed values come from EINV2 on held-out BP01 and SpatialScorer on BP02 and BP05, and SRP, SpatialScorer, ACCDOA, EINV2, and AudioANN-locfirst show comparable LOO transfer, all of them scoring visible candidates directly or routing attribution through localization. These values are descriptive: the paired tests in Table~\ref{tab:transfer_significance} do not resolve the order of these methods. Uniform-fold and sample-weighted LOO give 55.4/60.8\% for ACCDOA, 55.2/60.2\% for EINV2, 53.0/57.5\% for AudioANN-locfirst, and 53.5/52.9\% for SRP (Table~\ref{tab:loo_weighted}); removing any single bird keeps the AudioANN-locfirst mean between 50.8\% and 56.6\%, and excluding any complete experiment keeps it at or above 44.7\% while AudioANN-direct stays at 0.0\%. The BP05 pattern is not reducible to class imbalance alone.

\paragraph{Full attribution matrix.}
Table~\ref{tab:main_matrix_full} gives the full attribution matrix (24 methods plus the withheld on-body reference) corresponding to the compact main-text Table~\ref{tab:main_matrix}.

\begin{table*}[h!]
\centering
\caption{\textbf{Full vocal attribution matrix across benchmark regimes.} All methods solve the same task on the released events: given a synchronized multimodal window, pick the vocalizing bird from the set of candidates visible at that moment. Entries are raw attribution accuracy (\%). \textbf{Session} tests new recordings of known birds. \textbf{Held-out BP01/BP02/BP05} rotate the test experiment and measure transfer across experiments and individuals. \textbf{LOO mean} averages within-experiment leave-one-bird-out folds and isolates identity shift under a fixed sensing setup. Learned rows report mean $\pm$ standard deviation; deterministic rows are shown without $\pm$. The withheld on-body reference is included only as a ceiling.}
\label{tab:main_matrix_full}
\small
\setlength{\tabcolsep}{4pt}
\resizebox{\textwidth}{!}{%
\begin{tabular}{lccccc}
\toprule
Method & Session & Held-out BP01 & Held-out BP02 & Held-out BP05 & LOO mean \\
\midrule
Random & 28.3 $\pm$ 0.5 & 24.8 $\pm$ 0.2 & 33.3 $\pm$ 0.6 & 24.6 $\pm$ 0.3 & 27.0 $\pm$ 4.6 \\
LoudestMic & 31.1 & 18.9 & 39.5 & 26.3 & 27.3 $\pm$ 11.2 \\
SRP $\pm$20 & 51.0 & 52.4 & 62.2 & 35.8 & 53.5 $\pm$ 27.5 \\
\midrule
SELDnet & 81.8 $\pm$ 1.2 & 34.3 $\pm$ 0.4 & 38.6 $\pm$ 0.2 & 41.7 $\pm$ 8.6 & 0.0 $\pm$ 0.0 \\
SELDnet++ & \textbf{86.5 $\pm$ 0.4} & 39.8 $\pm$ 0.8 & 44.0 $\pm$ 1.2 & 36.9 $\pm$ 3.5 & 0.0 $\pm$ 0.0 \\
SELDnet++ open & 61.5 $\pm$ 0.4 & 34.6 $\pm$ 1.4 & 45.4 $\pm$ 0.0 & 29.2 $\pm$ 1.2 & 9.3 $\pm$ 4.8 \\
PairRank & 37.9 $\pm$ 1.7 & 32.7 $\pm$ 0.2 & 35.8 $\pm$ 1.1 & 29.7 $\pm$ 1.5 & 6.7 $\pm$ 5.0 \\
SpatialScorer & 73.0 $\pm$ 1.2 & 56.7 $\pm$ 1.6 & \textbf{69.1 $\pm$ 1.0} & \textbf{55.0 $\pm$ 1.2} & 49.5 $\pm$ 22.0 \\
SRP reranker & 45.1 $\pm$ 1.0 & 41.0 $\pm$ 1.2 & 51.6 $\pm$ 0.7 & 32.3 $\pm$ 0.5 & 38.5 $\pm$ 9.4 \\
NeuralSRP & 42.9 $\pm$ 0.9 & 26.0 $\pm$ 0.5 & 45.5 $\pm$ 0.8 & 49.2 $\pm$ 0.4 & 19.2 $\pm$ 13.8 \\
BeamFusion & 66.6 $\pm$ 2.5 & 44.9 $\pm$ 3.4 & 54.5 $\pm$ 3.3 & 34.1 $\pm$ 2.3 & 27.6 $\pm$ 12.2 \\
Contrastive & 72.7 $\pm$ 0.7 & 53.2 $\pm$ 1.2 & 66.4 $\pm$ 0.8 & 52.5 $\pm$ 1.9 & 40.4 $\pm$ 20.0 \\
Whisper candidate & 40.5 $\pm$ 0.8 & 27.6 $\pm$ 2.0 & 34.3 $\pm$ 0.2 & 32.2 $\pm$ 1.9 & 12.1 $\pm$ 6.0 \\
ACCDOA & 65.6 $\pm$ 0.2 & 58.1 $\pm$ 2.0 & 63.4 $\pm$ 1.3 & 38.6 $\pm$ 0.9 & \textbf{55.4 $\pm$ 21.8} \\
EINV2 & 65.8 $\pm$ 0.3 & \textbf{58.2 $\pm$ 1.9} & 66.6 $\pm$ 1.4 & 38.2 $\pm$ 0.3 & 55.2 $\pm$ 22.1 \\
CST-former & 49.8 $\pm$ 0.6 & 28.4 $\pm$ 0.6 & 35.8 $\pm$ 0.7 & 25.8 $\pm$ 0.9 & 15.0 $\pm$ 5.6 \\
\midrule
AudioANN-direct & 72.1 $\pm$ 3.1 & 35.0 $\pm$ 0.4 & 42.4 $\pm$ 0.5 & 28.8 $\pm$ 12.0 & 0.0 $\pm$ 0.0 \\
AudioANN-locfirst & 61.8 $\pm$ 0.6 & 56.3 $\pm$ 1.1 & 63.0 $\pm$ 0.2 & 36.3 $\pm$ 0.5 & 53.0 $\pm$ 19.2 \\
AudioANN-hybrid & 68.0 $\pm$ 3.8 & 35.8 $\pm$ 1.3 & 39.5 $\pm$ 2.7 & 22.8 $\pm$ 10.7 & 0.0 $\pm$ 0.0 \\
RadioANN-direct & 28.5 $\pm$ 0.4 & 29.2 $\pm$ 0.9 & 39.7 $\pm$ 0.3 & 25.4 $\pm$ 0.5 & 19.7 $\pm$ 9.5 \\
RadioANN-locfirst & 37.6 $\pm$ 0.1 & 32.4 $\pm$ 0.6 & 38.8 $\pm$ 0.2 & 29.3 $\pm$ 0.2 & 28.4 $\pm$ 11.6 \\
RadioANN-hybrid & 33.7 $\pm$ 0.4 & 30.5 $\pm$ 0.6 & 41.5 $\pm$ 0.6 & 28.3 $\pm$ 0.9 & 23.6 $\pm$ 11.8 \\
PoseMotion & 42.5 $\pm$ 0.2 & 35.8 $\pm$ 0.7 & 43.4 $\pm$ 0.3 & 28.8 $\pm$ 0.7 & 17.0 $\pm$ 7.3 \\
VideoCandidate & 44.6 $\pm$ 0.8 & 25.0 $\pm$ 0.5 & 37.2 $\pm$ 0.1 & 33.3 $\pm$ 1.1 & 26.9 $\pm$ 10.3 \\
\midrule
\multicolumn{6}{@{}l}{\emph{Ceiling / sanity-check row}} \\
On-body ref. & 99.2 $\pm$ 0.1 & 91.9 $\pm$ 1.0 & 98.0 $\pm$ 0.6 & 96.6 $\pm$ 0.2 & 91.3 $\pm$ 7.6 \\
\bottomrule
\end{tabular}%
}
\end{table*}

\subsection{Modality Interventions: Detailed Analysis}
\label{sec:appendix_intervention_detail}

This subsection expands the main-text intervention paragraph (§\ref{sec:diagnostics}).

\paragraph{CST-former and further interpretation.}
Under the short-window feature summary used in Table~\ref{tab:main_matrix}, radio is not intrinsically uninformative but is largely redundant once candidate geometry is known. CST-former shows small and sometimes counterintuitive deltas throughout, consistent with a model that does not strongly use any single candidate-side modality.

\paragraph{Transfer-rank significance.}
Tables~\ref{tab:transfer_attr_summary}--\ref{tab:transfer_significance} complete the picture: the strongest attribution rows form a compact spatially grounded tier under mean transfer rank, with SpatialScorer, SRP, ACCDOA, EINV2, and AudioANN-locfirst statistically indistinguishable, while CST-former and RadioANN-locfirst remain significantly weaker.

\paragraph{Position-noise robustness.}
A test-time perturbation of candidate 3D position with Gaussian noise at $\sigma \in \{1, 5, 10\}$~cm (Table~\ref{tab:geom_robust}) confirms the same split: SpatialScorer, Contrastive, BeamFusion, ACCDOA, and EINV2 degrade gracefully with noise magnitude, while NeuralSRP and SELDnet++ open are effectively flat under position noise, meaning they do not route attribution through candidate geometry. Head-orientation perturbation moved every cell by less than $0.1$ point and so does not contribute to attribution in any of these rows.

\paragraph{Full intervention table.}
Table~\ref{tab:shuffle_full} gives the full modality-intervention matrix across all four evaluation splits, with accuracy / macro-F1 / localization-error entries; the main-text Table~\ref{tab:main_matrix} shows only the LOO-mean accuracy deltas, including the trivially-zero deltas of the slot-collapse rows (SELDnet++, AudioANN-direct) for completeness.

\begin{table*}[h!]
\centering
\caption{\textbf{Full modality interventions.} Base rows report \textit{accuracy / legacy fold-wise macro-F1 / localization error in cm}; delta rows report changes from base in \textit{points / points / cm}. The macro-F1 entries are retained only as intervention diagnostics within the same split and should not be read as the LOO class-balanced benchmark summary. Positive localization deltas indicate worse error. Interventions are evaluated on a single seed-42 checkpoint, so base accuracies may differ slightly from the multi-seed means reported in Table~\ref{tab:main_matrix_full} while remaining within the reported seed variance.}
\label{tab:shuffle_full}
\small
\setlength{\tabcolsep}{4pt}
\resizebox{\textwidth}{!}{%
\begin{tabular}{llcccc}
\toprule
Split & Condition & ACCDOA & EINV2 & SpatialScorer & CST-former \\
\midrule
\multirow{4}{*}{Held-out BP01}
& \cellcolor{gray!12}Base & 58.4/56.7/23.2 & 57.9/56.1/24.2 & 58.5/57.1/17.0 & 28.4/26.6/22.0 \\
& \cellcolor{red!10}$\Delta$ zero audio & -27.9/-27.2/+9.2 & -24.5/-23.8/+3.8 & -35.9/-35.4/+15.3 & +4.1/+4.2/+5.9 \\
& \cellcolor{blue!10}$\Delta$ shuffle geometry & -33.7/-32.5/+0.0 & -33.1/-31.9/+0.0 & -33.5/-32.5/+0.0 & -0.7/-0.7/+0.0 \\
& \cellcolor{green!10}$\Delta$ zero radio & +0.0/+0.0/+0.0 & +0.0/+0.0/+0.0 & +1.2/+1.4/+0.0 & +5.1/+5.8/+0.0 \\
\midrule
\multirow{4}{*}{Held-out BP02}
& \cellcolor{gray!12}Base & 61.2/59.4/22.7 & 66.6/65.2/21.8 & 66.3/65.7/19.8 & 36.3/35.3/25.8 \\
& \cellcolor{red!10}$\Delta$ zero audio & -25.2/-24.3/+10.2 & -29.8/-29.7/+11.5 & -27.1/-28.8/+14.2 & -2.7/-2.3/+3.8 \\
& \cellcolor{blue!10}$\Delta$ shuffle geometry & -27.5/-26.8/+0.0 & -32.7/-32.5/+0.0 & -31.9/-32.5/+0.0 & -1.3/-1.4/+0.0 \\
& \cellcolor{green!10}$\Delta$ zero radio & +0.0/+0.0/+0.0 & +0.0/+0.0/+0.0 & -0.8/-1.1/+0.0 & -0.1/+0.3/+0.0 \\
\midrule
\multirow{4}{*}{Held-out BP05}
& \cellcolor{gray!12}Base & 37.8/26.9/26.9 & 38.1/27.2/26.7 & 53.4/32.3/23.9 & 25.7/17.9/26.2 \\
& \cellcolor{red!10}$\Delta$ zero audio & -9.1/-7.1/+5.0 & -6.6/-6.2/+2.9 & -3.2/-5.7/+8.0 & -1.3/+0.3/+3.7 \\
& \cellcolor{blue!10}$\Delta$ shuffle geometry & -11.9/-9.5/+0.0 & -12.3/-10.1/+0.0 & -16.8/-10.5/+0.0 & -3.9/-2.7/+0.0 \\
& \cellcolor{green!10}$\Delta$ zero radio & +0.0/+0.0/+0.0 & +0.0/+0.0/+0.0 & -14.8/-5.1/+0.0 & +6.0/+3.0/+0.0 \\
\midrule
\multirow{4}{*}{LOO mean}
& \cellcolor{gray!12}Base & 56.1/19.4/23.7 & 54.4/18.9/25.0 & 48.7/17.2/19.8 & 15.1/7.1/23.0 \\
& \cellcolor{red!10}$\Delta$ zero audio & -27.4/-7.3/+8.2 & -26.4/-7.0/+6.9 & -26.2/-7.9/+11.0 & -2.3/-1.3/+3.1 \\
& \cellcolor{blue!10}$\Delta$ shuffle geometry & -29.0/-7.6/+0.0 & -26.9/-6.8/+0.0 & -25.2/-7.1/+0.0 & -0.2/+0.0/+0.0 \\
& \cellcolor{green!10}$\Delta$ zero radio & +0.0/+0.0/+0.0 & +0.0/+0.0/+0.0 & +6.3/+1.7/+0.0 & +10.8/+4.1/+0.0 \\
\bottomrule
\end{tabular}%
}
\end{table*}

\subsection{Per-Bird Leave-One-Bird-Out Results}

Table~\ref{tab:loo_perbird} reports fold-level LOO attribution accuracy for representative methods. Fold difficulty is heterogeneous and not reducible to class frequency alone: the smallest BP05 bird (56 test events) is not uniformly the hardest fold, while the largest BP05 fold can still break otherwise strong models. The withheld on-body reference remains strong on all BP05 folds, which supports the interpretation that difficult folds reflect modeling challenges rather than invalid labels.

\begin{table*}[!htb]
\centering
\caption{\textbf{Per-bird leave-one-bird-out accuracy (\%).} Birds are rows and representative completed methods are columns. \textbf{Test $n$} is the number of held-out test events in that fold. For learned methods, entries report mean $\pm$ standard deviation across seeds within that held-out bird fold.}
\label{tab:loo_perbird}
\tiny
\resizebox{\textwidth}{!}{%
\begin{tabular}{lccccccccc}
\toprule
Bird & Test $n$ & SRP $\pm$20 & SpatialScorer & ACCDOA & EINV2 & Contrastive & VideoCandidate & On-body ref. & AudioANN-locfirst \\
\midrule
BP01-0 & 3963 & 75.7 & 61.9 $\pm$ 3.3 & 67.4 $\pm$ 0.5 & 67.8 $\pm$ 0.7 & 46.7 $\pm$ 1.3 & 30.6 $\pm$ 0.4 & 98.6 $\pm$ 0.2 & 63.3 $\pm$ 0.4 \\
BP01-1 & 2277 & 31.8 & 60.9 $\pm$ 2.6 & 65.0 $\pm$ 1.6 & 66.3 $\pm$ 0.8 & 58.7 $\pm$ 2.7 & 46.9 $\pm$ 0.1 & 93.8 $\pm$ 0.8 & 56.8 $\pm$ 0.8 \\
BP01-2 & 4244 & 71.2 & 62.9 $\pm$ 0.6 & 73.5 $\pm$ 0.4 & 73.9 $\pm$ 0.8 & 63.4 $\pm$ 1.3 & 26.5 $\pm$ 0.2 & 86.8 $\pm$ 2.5 & 72.5 $\pm$ 0.3 \\
BP01-3 & 5735 & 30.5 & \textbf{71.1 $\pm$ 1.2} & 68.6 $\pm$ 1.2 & 69.1 $\pm$ 0.8 & 55.1 $\pm$ 3.2 & 38.2 $\pm$ 1.0 & 98.1 $\pm$ 0.5 & 69.3 $\pm$ 1.1 \\
BP02-0 & 2065 & \textbf{88.8} & 40.7 $\pm$ 1.1 & 71.1 $\pm$ 1.0 & 71.0 $\pm$ 1.6 & 35.8 $\pm$ 1.2 & 14.2 $\pm$ 0.3 & 97.2 $\pm$ 0.4 & 62.8 $\pm$ 1.7 \\
BP02-1 & 4589 & 61.0 & 63.8 $\pm$ 1.1 & 62.3 $\pm$ 0.7 & 60.7 $\pm$ 0.8 & 28.9 $\pm$ 0.5 & 27.8 $\pm$ 1.5 & 79.1 $\pm$ 9.5 & 59.6 $\pm$ 2.1 \\
BP02-2 & 4923 & 52.1 & 64.3 $\pm$ 1.7 & \textbf{73.0 $\pm$ 0.9} & 71.4 $\pm$ 3.5 & 59.1 $\pm$ 3.2 & 30.3 $\pm$ 0.9 & 97.1 $\pm$ 0.4 & 65.5 $\pm$ 4.3 \\
BP05-0 & 2140 & \textbf{85.6} & 30.9 $\pm$ 2.0 & 25.4 $\pm$ 2.7 & 22.8 $\pm$ 2.9 & 23.7 $\pm$ 0.9 & 11.7 $\pm$ 0.7 & 90.4 $\pm$ 2.6 & 21.5 $\pm$ 0.4 \\
BP05-2 & 56 & \textbf{32.1} & 0.0 $\pm$ 0.0 & 26.8 $\pm$ 3.1 & 30.4 $\pm$ 1.8 & 0.0 $\pm$ 0.0 & 26.8 $\pm$ 4.7 & 94.0 $\pm$ 2.7 & \textbf{38.1 $\pm$ 2.2} \\
BP05-3 & 3633 & 6.6 & \textbf{39.0 $\pm$ 1.4} & 20.5 $\pm$ 0.4 & 18.5 $\pm$ 0.9 & 33.0 $\pm$ 2.0 & 16.4 $\pm$ 0.7 & 78.2 $\pm$ 2.3 & 20.6 $\pm$ 0.8 \\
\bottomrule
\end{tabular}}
\end{table*}

\paragraph{What predicts fold difficulty?}
To separate sample-count and geometry effects, Table~\ref{tab:loo_fold_drivers} correlates per-fold LOO accuracy across the 10 held-out-bird folds with (i) test-set size and (ii) the held-out bird's mean nearest-neighbor distance, derived from Table~\ref{tab:pairwise}. The picture is mixed rather than one-factor. For learned candidate models, fold size is often the stronger coarse predictor, but it is not sufficient: BP05-0 and BP05-3 remain difficult despite 2{,}140 and 3{,}633 test events, and the on-body reference stays high on both. For SRP, nearest-neighbor separation is the clearer predictor, consistent with its direct dependence on acoustic ambiguity. These summaries support the main interpretation in the paper: BP05 imbalance increases variance, but the transfer ordering cannot be reduced to sample count alone.

\begin{table}[!htb]
\centering
\caption{\textbf{Simple predictors of LOO fold difficulty.} Spearman correlation $\rho$ across the 10 held-out-bird folds between fold-wise LOO accuracy, test-set size, and the held-out bird's mean nearest-neighbor distance.}
\label{tab:loo_fold_drivers}
\small
\setlength{\tabcolsep}{5pt}
\begin{tabular}{lcc}
\toprule
Method & $\rho(\mathrm{acc}, n_{\mathrm{test}})$ & $\rho(\mathrm{acc}, d_{\mathrm{NN}})$ \\
\midrule
SRP $\pm 20$ & -0.30 & \phantom{-}0.45 \\
SpatialScorer & \phantom{-}0.94 & -0.23 \\
ACCDOA & \phantom{-}0.44 & -0.11 \\
EINV2 & \phantom{-}0.44 & -0.11 \\
AudioANN-locfirst & \phantom{-}0.65 & -0.27 \\
On-body ref. & -0.06 & -0.20 \\
\bottomrule
\end{tabular}
\end{table}

\subsection{Sample-Weighted LOO and Per-Bird Recall}
\label{sec:appendix_weighted_loo}

The main-text LOO column averages the 10 held-out-bird folds uniformly, giving equal weight to each held-out vocalizer. Because each LOO test fold contains positive calls from exactly one physical bird, this uniform mean is the per-bird recall average across the 10 released individuals. This is the cleaner class-balanced summary than fold-wise macro-F1 over local slot indices, which assigns zero F1 to absent classes and can make the near-perfect on-body reference appear weak. To show that the benchmark conclusions do not depend on uniform fold weighting, Table~\ref{tab:loo_weighted} also reports sample-weighted LOO averages, where folds are weighted by their test-set size (Table~\ref{tab:loo_sizes}). The sample-weighted column explicitly down-weights the very small BP05 Bird~2 fold (56 test events), and the same spatially grounded top tier remains visible.

\begin{table}[!htb]
\centering
\caption{\textbf{LOO accuracy / per-bird recall, uniform vs.\ sample-weighted across the 10 held-out-bird folds.} Uniform columns match Table~\ref{tab:main_matrix} and are also the per-bird recall average. Weighted columns use the test-set sizes in Table~\ref{tab:loo_sizes}. Both columns average seeds within each fold before aggregating.}
\label{tab:loo_weighted}
\small
\setlength{\tabcolsep}{5pt}
\begin{tabular}{lcc}
\toprule
Method & Uniform / per-bird recall & Sample-weighted \\
\midrule
Random              & 27.0 & 27.9 \\
SRP $\pm 20$        & 53.5 & 52.9 \\
\midrule
SELDnet             & \phantom{0}0.0 & \phantom{0}0.0 \\
SpatialScorer       & 49.5 & 58.3 \\
NeuralSRP           & 19.2 & 22.8 \\
BeamFusion          & 27.6 & 30.5 \\
Contrastive         & 40.4 & 46.8 \\
Whisper candidate   & 12.1 & 13.3 \\
ACCDOA              & \textbf{55.4} & \textbf{60.8} \\
EINV2               & 55.2 & 60.2 \\
CST-former          & 15.0 & 16.3 \\
PairRank            & \phantom{0}6.7 & \phantom{0}6.8 \\
\midrule
AudioANN-direct     & \phantom{0}0.0 & \phantom{0}0.0 \\
AudioANN-locfirst   & 53.0 & 57.5 \\
AudioANN-hybrid     & \phantom{0}0.0 & \phantom{0}0.0 \\
RadioANN-direct     & 19.7 & 20.1 \\
RadioANN-locfirst   & 28.4 & 30.5 \\
RadioANN-hybrid     & 23.6 & 25.4 \\
PoseMotion          & 17.0 & 19.4 \\
VideoCandidate      & 26.9 & 28.3 \\
\midrule
On-body ref.        & 91.3 & 90.9 \\
\bottomrule
\end{tabular}
\end{table}

\subsection{Localization Results}

Table~\ref{tab:localization_matrix} reports the full localization companion results referenced briefly in Section~\ref{sec:experiments}. Free 3D localization and candidate-conditioned localization are kept separate because they correspond to different formulations of the task; the withheld on-body reference is meaningful only as a ceiling for the candidate-conditioned setting.

\begin{table*}[!htb]
\centering
\caption{\textbf{Localization results} (mean error in cm, lower is better). Free 3D localization and candidate-conditioned localization are reported separately. The withheld on-body reference is a ceiling only for the candidate-conditioned setting.}
\label{tab:localization_matrix}
\small
\setlength{\tabcolsep}{4pt}
\resizebox{\textwidth}{!}{%
\begin{tabular}{lccccc}
\toprule
Method & Session & Held-out BP01 & Held-out BP02 & Held-out BP05 & LOO mean \\
\midrule
\multicolumn{6}{@{}l}{\emph{Free 3D localization}} \\
ACCDOA & 22.7 $\pm$ 0.4 & 24.0 $\pm$ 1.4 & 22.2 $\pm$ 0.4 & 28.6 $\pm$ 0.8 & 25.1 $\pm$ 7.5 \\
EINV2 & \textbf{20.7 $\pm$ 0.6} & 23.9 $\pm$ 0.8 & \textbf{21.7 $\pm$ 0.4} & 27.1 $\pm$ 0.7 & 25.0 $\pm$ 9.6 \\
CST-former & 21.8 $\pm$ 0.4 & \textbf{22.8 $\pm$ 0.7} & 25.8 $\pm$ 0.4 & \textbf{26.9 $\pm$ 0.5} & \textbf{23.2 $\pm$ 5.0} \\
AudioANN-locfirst & 21.7 $\pm$ 0.7 & 25.0 $\pm$ 0.4 & 27.2 $\pm$ 2.7 & 27.2 $\pm$ 0.7 & 24.6 $\pm$ 8.0 \\
RadioANN-locfirst & 41.1 $\pm$ 0.6 & 27.2 $\pm$ 0.1 & 49.4 $\pm$ 1.8 & 49.1 $\pm$ 2.2 & 34.9 $\pm$ 6.7 \\
\midrule
\multicolumn{6}{@{}l}{\emph{Candidate-conditioned localization}} \\
Random candidate & 27.5 & 26.9 & 27.1 & 28.4 & 27.1 $\pm$ 1.3 \\
SELDnet & 25.6 $\pm$ 0.1 & 25.5 $\pm$ 0.5 & 30.5 $\pm$ 0.3 & 28.8 $\pm$ 0.0 & 26.1 $\pm$ 3.9 \\
PairRank & 21.8 $\pm$ 0.5 & 21.6 $\pm$ 0.2 & 24.4 $\pm$ 1.4 & 23.4 $\pm$ 0.9 & 32.8 $\pm$ 6.6 \\
SpatialScorer & 18.2 $\pm$ 0.1 & \textbf{17.4 $\pm$ 0.1} & \textbf{19.2 $\pm$ 0.2} & 24.2 $\pm$ 0.3 & \textbf{19.7 $\pm$ 5.5} \\
NeuralSRP & 21.3 $\pm$ 0.2 & 22.6 $\pm$ 0.5 & \textbf{20.5 $\pm$ 0.2} & 22.2 $\pm$ 0.0 & 27.0 $\pm$ 4.4 \\
BeamFusion & 11.7 $\pm$ 0.7 & 17.1 $\pm$ 1.6 & 17.1 $\pm$ 1.1 & \textbf{20.5 $\pm$ 0.6} & 23.7 $\pm$ 6.7 \\
Contrastive & 19.1 $\pm$ 0.3 & 19.4 $\pm$ 0.4 & 22.3 $\pm$ 0.5 & 26.0 $\pm$ 0.4 & 23.0 $\pm$ 6.6 \\
Whisper candidate & 21.9 $\pm$ 0.1 & 22.2 $\pm$ 0.2 & 24.1 $\pm$ 0.1 & 22.9 $\pm$ 0.2 & 26.7 $\pm$ 5.2 \\
PoseMotion & 20.5 $\pm$ 0.1 & 21.0 $\pm$ 0.1 & 21.9 $\pm$ 0.1 & 22.0 $\pm$ 0.1 & 25.4 $\pm$ 4.6 \\
VideoCandidate & 19.9 $\pm$ 0.1 & 22.5 $\pm$ 0.2 & 23.6 $\pm$ 0.2 & 21.9 $\pm$ 0.1 & 23.0 $\pm$ 4.7 \\
\midrule
\multicolumn{6}{@{}l}{\emph{Candidate-conditioned ceiling}} \\
On-body ref. & 0.3 $\pm$ 0.0 & 2.9 $\pm$ 0.4 & 0.8 $\pm$ 0.2 & 1.4 $\pm$ 0.1 & 3.3 $\pm$ 3.1 \\
\bottomrule
\end{tabular}}
\end{table*}

\subsection{Global Transfer Summaries}

Following \citet{demsar2006statistical}, we summarize the benchmark not only by aggregated regime means, but also by task-wise ranks over the 13 transfer tasks: the three held-out-experiment folds and the ten leave-one-bird-out folds. Table~\ref{tab:transfer_attr_summary} reports the mean transfer accuracy for every reported attribution method, a percentile-bootstrap 95\% interval for that mean obtained by resampling transfer tasks with replacement~\citep{efron1993introduction}, the average task rank (lower is better), and the number of fold wins. The on-body reference is shown separately as a ceiling and is not included in the rank competition. Table~\ref{tab:transfer_loc_summary} gives the same summaries for localization, with free and candidate-conditioned formulations kept separate.

\begin{table*}[!htb]
\centering
\begin{minipage}[t]{0.48\textwidth}
\centering
\captionof{table}{\textbf{Global transfer attribution summary.} Mean accuracy over 13 transfer tasks (3 held-out + 10 LOO). Rank and wins computed task-wise.}
\label{tab:transfer_attr_summary}
\scriptsize
\setlength{\tabcolsep}{3pt}
\begin{tabular}{lcccc}
\toprule
Method & Acc. & 95\% CI & Rank & Wins \\
\midrule
Random & 27.2 & [24.9, 29.5] & 12.9 & 0 \\
SRP $\pm 20$ & 52.7 & [40.0, 65.5] & 5.9 & 3 \\
SELDnet & 8.8 & [0.0, 18.0] & 20.2 & 0 \\
PairRank & 12.7 & [6.7, 19.6] & 16.4 & 0 \\
SpatialScorer & 52.0 & [40.7, 61.2] & 4.1 & 5 \\
SRP reranker & 39.2 & [34.4, 43.8] & 6.9 & 0 \\
NeuralSRP & 24.1 & [15.7, 32.2] & 11.7 & 0 \\
BeamFusion & 31.5 & [24.4, 38.3] & 9.6 & 0 \\
Contrastive & 44.4 & [33.7, 53.7] & 6.4 & 0 \\
Whisper cand. & 16.5 & [11.4, 22.0] & 15.9 & 0 \\
ACCDOA & 54.9 & [44.2, 64.5] & 3.6 & 1 \\
EINV2 & 55.0 & [43.9, 64.6] & 3.8 & 3 \\
CST-former & 18.5 & [14.2, 23.0] & 15.2 & 0 \\
AudioANN-dir & 8.2 & [0.0, 17.3] & 18.1 & 0 \\
AudioANN-loc & 52.7 & [42.9, 61.5] & 4.7 & 1 \\
AudioANN-hyb & 7.5 & [0.0, 16.1] & 18.7 & 0 \\
RadioANN-dir & 22.4 & [17.1, 27.6] & 13.1 & 0 \\
RadioANN-loc & 29.6 & [24.1, 35.0] & 10.2 & 0 \\
RadioANN-hyb & 25.9 & [19.8, 31.7] & 11.8 & 0 \\
PoseMotion & 21.4 & [15.8, 27.5] & 13.5 & 0 \\
VideoCandidate & 28.1 & [22.8, 33.2] & 12.4 & 0 \\
\midrule
On-body ref. & 92.7 & [88.4, 96.1] & -- & -- \\
\bottomrule
\end{tabular}
\end{minipage}%
\hfill
\begin{minipage}[t]{0.48\textwidth}
\centering
\captionof{table}{\textbf{Global transfer localization summary.} Mean error (cm) over the same 13 tasks. Rank within formulation only.}
\label{tab:transfer_loc_summary}
\scriptsize
\setlength{\tabcolsep}{3pt}
\begin{tabular}{lcccc}
\toprule
Method & Error & 95\% CI & Rank & Wins \\
\midrule
\multicolumn{5}{@{}l}{\emph{Free 3D localization}} \\
ACCDOA & 25.1 & [21.8, 28.8] & 3.0 & 0 \\
EINV2 & 24.9 & [20.8, 29.6] & 2.5 & 4 \\
CST-former & 23.7 & [21.3, 26.0] & 2.3 & 6 \\
AudioANN-loc & 25.0 & [21.4, 28.9] & 2.7 & 3 \\
RadioANN-loc & 36.5 & [32.2, 41.0] & 4.5 & 0 \\
\midrule
\multicolumn{5}{@{}l}{\emph{Candidate-conditioned}} \\
SELDnet & 26.6 & [24.6, 28.4] & 7.5 & 0 \\
PairRank & 30.6 & [27.0, 34.5] & 8.4 & 0 \\
SpatialScorer & 19.9 & [17.4, 22.6] & 2.3 & 6 \\
NeuralSRP & 25.8 & [23.7, 28.4] & 5.7 & 1 \\
BeamFusion & 22.5 & [19.3, 25.9] & 3.3 & 4 \\
Contrastive & 22.9 & [19.9, 26.1] & 4.5 & 0 \\
Whisper cand. & 25.8 & [23.5, 28.5] & 6.0 & 0 \\
PoseMotion & 24.5 & [22.5, 27.2] & 4.6 & 1 \\
VideoCandidate & 22.9 & [20.8, 25.3] & 3.9 & 1 \\
\midrule
On-body ref. & 2.8 & [1.4, 4.5] & -- & -- \\
\bottomrule
\end{tabular}
\end{minipage}
\end{table*}

\subsection{Selected Paired Transfer Tests}

For focused comparative claims among the strongest transfer methods, Table~\ref{tab:transfer_significance} reports paired two-sided Wilcoxon signed-rank tests across the same 13 transfer tasks. We additionally report Holm-adjusted $p$ values~\citep{holm1979simple} for the family of accuracy comparisons in the table.

\begin{table}[!htb]
\centering
\caption{\textbf{Selected paired transfer significance tests for attribution accuracy.} Positive $\Delta$acc means that the row method improves mean fold-wise transfer accuracy over the column method.}
\label{tab:transfer_significance}
\small
\setlength{\tabcolsep}{4pt}
\begin{tabular}{lrrr}
\toprule
Comparison & $\Delta$acc & $p$ & $p_{\mathrm{Holm}}$ \\
\midrule
SpatialScorer $-$ SRP $\pm 20$ & -0.007 & 0.94604 & 1.00000 \\
ACCDOA $-$ SRP $\pm 20$ & +0.022 & 0.45483 & 1.00000 \\
EINV2 $-$ SRP $\pm 20$ & +0.022 & 0.37573 & 1.00000 \\
AudioANN-locfirst $-$ SRP $\pm 20$ & -0.000 & 0.49731 & 1.00000 \\
CST-former $-$ SRP $\pm 20$ & -0.343 & 0.00049 & 0.00391 \\
RadioANN-locfirst $-$ SRP $\pm 20$ & -0.231 & 0.00464 & 0.03247 \\
ACCDOA $-$ SpatialScorer & +0.029 & 0.63550 & 1.00000 \\
EINV2 $-$ SpatialScorer & +0.030 & 0.63550 & 1.00000 \\
\bottomrule
\end{tabular}
\end{table}

\subsection{Geometry robustness under test-time perturbation}
\label{sec:appendix_geom_robust}

To measure how much the geometry-consuming rows rely on clean candidate state, we perturb candidate 3D positions with Gaussian noise at $\sigma \in \{1, 5, 10\}$~cm and head orientations by axis-angle rotations with Gaussian angle $\sigma \in \{5, 15, 30\}^\circ$ at test time only, applied to the best-validation checkpoint of each model. Table~\ref{tab:geom_robust} reports mean attribution accuracy across seeds (and LOO folds, for the LOO row). Session is averaged over 3 seeds, Heldout over 9 (3 experiments $\times$ 3 seeds), and LOO over 30 (10 birds $\times$ 3 seeds). Orientation perturbation did not change attribution accuracy by more than 0.1 points on any model and split, so head direction is not contributing to attribution in this baseline suite under short windows; we omit orientation columns from the table for brevity. Under position noise, SpatialScorer, Contrastive, BeamFusion, ACCDOA, and EINV2 degrade gracefully (a few points at $\sigma{=}5$~cm, up to $\sim 10$ points at $\sigma{=}10$~cm on session, less on transfer splits), NeuralSRP is insensitive to position noise after training, and SELDnet++ open (trained at the reduced 8-epoch budget used for this sweep) is also effectively flat under position perturbation, consistent with a model that does not route attribution through candidate geometry.

\begin{table*}[!htb]
\centering
\caption{\textbf{Mean attribution accuracy under test-time perturbation of candidate 3D position.} Rows are grouped by evaluation regime. $\Delta_{10}$ is the drop at $\sigma{=}10$~cm relative to the clean column. Orientation perturbations (not shown) moved every cell by $<0.1$ pt.}
\label{tab:geom_robust}
\small
\setlength{\tabcolsep}{4pt}
\begin{tabular}{llccccc}
\toprule
Regime & Method & Clean & $\sigma{=}1$~cm & $\sigma{=}5$~cm & $\sigma{=}10$~cm & $\Delta_{10}$ \\
\midrule
\multirow{7}{*}{Session}
 & SpatialScorer     & 71.8 & 71.5 & 68.3 & 62.1 & -9.7 \\
 & Contrastive       & 72.0 & 71.9 & 69.9 & 64.7 & -7.3 \\
 & BeamFusion        & 67.9 & 67.3 & 63.0 & 55.1 & -12.8 \\
 & ACCDOA            & 64.5 & 64.4 & 61.9 & 56.2 & -8.3 \\
 & EINV2             & 65.2 & 65.1 & 62.6 & 56.5 & -8.7 \\
 & NeuralSRP         & 43.8 & 43.8 & 43.7 & 43.6 & -0.2 \\
 & SELDnet++ open    & 47.7 & 47.7 & 47.6 & 47.6 & -0.1 \\
\midrule
\multirow{7}{*}{Heldout (avg.~BP01/02/05)}
 & SpatialScorer     & 54.9 & 54.8 & 53.0 & 49.3 & -5.6 \\
 & Contrastive       & 59.8 & 59.8 & 58.3 & 54.1 & -5.7 \\
 & BeamFusion        & 45.5 & 45.3 & 42.4 & 38.1 & -7.4 \\
 & ACCDOA            & 52.1 & 51.9 & 49.9 & 45.5 & -6.6 \\
 & EINV2             & 52.2 & 52.0 & 50.1 & 45.5 & -6.7 \\
 & NeuralSRP         & 40.1 & 40.1 & 40.1 & 40.1 & +0.0 \\
 & SELDnet++ open    & 36.3 & 36.3 & 36.3 & 36.3 & +0.0 \\
\midrule
\multirow{7}{*}{LOO (avg.~10 birds)}
 & SpatialScorer     & 47.0 & 46.9 & 44.4 & 40.3 & -6.7 \\
 & Contrastive       & 41.6 & 41.6 & 40.4 & 37.3 & -4.3 \\
 & BeamFusion        & 28.6 & 28.3 & 26.0 & 22.8 & -5.8 \\
 & ACCDOA            & 54.4 & 54.4 & 52.3 & 47.6 & -6.8 \\
 & EINV2             & 54.4 & 54.3 & 52.6 & 47.5 & -6.9 \\
 & NeuralSRP         & 18.5 & 18.5 & 18.5 & 18.5 & +0.0 \\
 & SELDnet++ open    & 10.3 & 10.3 & 10.3 & 10.3 & +0.1 \\
\bottomrule
\end{tabular}
\end{table*}

\subsection{Attribution and Inter-Candidate Distance}
\label{app:distance}
We stratified the three-seed LOO evaluation, pooling test events across all ten held-out-bird folds, by the distance from the vocalizer to its nearest competing candidate (Table~\ref{tab:distance}). Accuracy increases as the nearest competitor moves farther away: ACCDOA rises by 26.4 points from the closest to the farthest bin, while random attribution changes by 4.8 points. We fit a logistic slope of correctness against standardized $\log_{10}$ nearest-competitor distance separately for each seed and fold, average the seed-level slopes within each fold, and apply a two-sided Wilcoxon signed-rank test to the ten fold-level slopes. Eight of ten ACCDOA slopes are positive ($p=0.0098$); Holm correction across AudioANN-locfirst, ACCDOA, SpatialScorer, and SRP gives $p=0.0293$.

\begin{table}[h]
\centering
\small
\caption{\textbf{LOO attribution accuracy (\%) by distance from the vocalizer to its nearest competing candidate.}}
\label{tab:distance}
\begin{tabular}{lrrr}
\toprule
Nearest competitor & ACCDOA & Random & Withheld on-body reference \\
\midrule
$<10$ cm & 46.6 & 26.3 & 91.2 \\
10--20 cm & 54.2 & 26.7 & 91.1 \\
20--30 cm & 61.0 & 27.8 & 91.2 \\
30--50 cm & 67.3 & 28.9 & 90.8 \\
$>50$ cm & 73.0 & 31.1 & 90.2 \\
\bottomrule
\end{tabular}
\end{table}

The array geometry continues to distinguish close candidates. Only 122 of 33{,}625 events (0.36\%) have a competitor whose expected delays are within one audio sample of the vocalizer on every microphone pair; all 122 occur below 10~cm and represent 3.66\% of that subset. A fixed GCC-PHAT decoder, which compares the observed microphone-pair correlations with the delays predicted from each candidate position and uses only the five-channel waveform, microphone geometry, and candidate positions, reaches 36.6\% below 10~cm against 26.3\% random attribution, and 38.4\% overall against 27.9\%; it remains above chance in every separation bin. Adding isotropic Gaussian noise to the verified vocalizer position and selecting the nearest candidate gives 91.0--98.9\% attribution across the ten LOO folds at $\sigma=50$~mm, 74.9--91.7\% at $\sigma=100$~mm, and 62.6--83.0\% at $\sigma=150$~mm.

\subsection{Microphone Ablations}
\label{app:mics}
In a matched single-checkpoint-per-fold LOO ablation, SpatialScorer falls from 51.2\% with five microphones to 45.2\%, 38.5\%, and 26.0\% with four, three, and zero microphones. The 51.2\% value is the intact baseline under this ablation protocol; Table~\ref{tab:main_matrix} reports the separate three-seed LOO summary of $49.5\pm22.0$\%. On session-disjoint evaluation, permuting the channel-to-microphone correspondence reduces ACCDOA from 64.3\% to 38.8\%. Across the ten LOO folds, classical SRP ranges from 88.8\% on BP02-0 to 6.6\% on BP05-3, showing sensitivity to room geometry and reverberation.

\subsection{Bioacoustic Front Ends and Clustering}
\label{app:frontends}
We replaced the frozen mel front end of one candidate-scoring head with AVES, BirdNET, and animal2vec, fixing the head, protocol, 60-epoch budget, and three seeds (Table~\ref{tab:frontends}). All front ends reach 74--75\% for familiar birds and 49--50\% held-out, but 22.8--24.1\% LOO against 27.0\% for random attribution. During LOO training the held-out bird can appear as a silent candidate but never as the vocalizer, which permits a negative prior; held-out-experiment birds are absent in both roles. Stronger frozen acoustic representations preserve familiar-bird information but do not by themselves establish which tracked candidate produced the vocalization.

\begin{table}[h]
\centering
\small
\caption{\textbf{Frozen audio front ends under a fixed candidate-scoring head} (accuracy, \%). The held-out column averages BP01, BP02, and BP05.}
\label{tab:frontends}
\begin{tabular}{lrrr}
\toprule
Front end & Session & Held-out exp.\ mean & LOO \\
\midrule
Random candidate & 28.3 & 27.6 & 27.0 \\
Mel spectrogram & $75.2 \pm 0.5$ & $49.1 \pm 0.5$ & $23.8 \pm 0.2$ \\
AVES (self-supervised) & $75.4 \pm 0.7$ & $49.3 \pm 1.0$ & $24.1 \pm 0.5$ \\
BirdNET (supervised) & $75.3 \pm 0.8$ & $50.0 \pm 0.4$ & $22.8 \pm 0.5$ \\
animal2vec (self-supervised) & $74.3 \pm 0.5$ & $50.0 \pm 1.4$ & $22.8 \pm 0.6$ \\
\bottomrule
\end{tabular}
\end{table}

For the unsupervised alternative, we clustered segmented events with $k$-means given the true number of birds, using a frozen encoder trained on the same cohort. NMI and ARI reach 0.61 and 0.58, but assigning $K$ clusters to $K$ tracked birds still requires resolving among $K!$ possible mappings.

\subsection{Orientation-Weighted LoudestMic}
\label{app:loudestmic_orient}
For each bird we compute the cosine alignment $c_i\in[-1,1]$ between its head-to-beak direction and the direction to the loudest microphone, multiply the bird-to-microphone distance by $1-0.5(c_i+1)/2$, and select the minimum; the weight was fixed at 0.5 without tuning. The orientation-weighted variant changes accuracy by less than 0.6 points, in directions that vary across regimes (session-disjoint 31.1\% $\to$ 30.8\%, held-out experiment 28.2\% $\to$ 28.7\%, LOO 27.3\% $\to$ 27.9\%), consistent with the small effect of orientation perturbation in Appendix~\ref{sec:appendix_geom_robust}. We retain the original LoudestMic as the simpler energy-plus-distance reference.

\clearpage
\section*{NeurIPS Paper Checklist}

\begin{enumerate}

\item {\bf Claims}
    \item[] Question: Do the main claims made in the abstract and introduction accurately reflect the paper's contributions and scope?
    \item[] Answer: \answerYes{}
    \item[] Justification: The abstract, introduction, and experiments sections consistently frame the paper as a multimodal benchmark for spatial vocal attribution with transfer-oriented evaluation, withheld on-body reference rows, and modality interventions; the main claims are scoped to that benchmark setting.
\item {\bf Limitations}
    \item[] Question: Does the paper discuss the limitations of the work performed by the authors?
    \item[] Answer: \answerYes{}
    \item[] Justification: Section~\ref{sec:limitations} discusses the benchmark scope explicitly, including the single sensing installation, current scale, and restriction to verified single-vocalizer events.
\item {\bf Theory assumptions and proofs}
    \item[] Question: For each theoretical result, does the paper provide the full set of assumptions and a complete (and correct) proof?
    \item[] Answer: \answerNA{}
    \item[] Justification: The paper does not present theoretical results requiring formal assumptions or proofs.
\item {\bf Open access to data and code}
    \item[] Question: Does the paper provide open access to the data and code, with sufficient instructions to faithfully reproduce the main experimental results, as described in supplemental material?
    \item[] Answer: \answerYes{}
    \item[] Justification: The introduction and the \emph{Ethics, Licensing, and Release} section provide public dataset and code URLs, and the appendix documents the preprocessing and training setup used for the benchmark.
\item {\bf Experimental setting/details}
    \item[] Question: Does the paper specify all the training and test details (e.g., data splits, hyperparameters, how they were chosen, type of optimizer) necessary to understand the results?
    \item[] Answer: \answerYes{}
    \item[] Justification: Sections~4--6 define the evaluation setup, and the appendix gives model-family-specific architectural and training details, including split construction and the preprocessing pipeline.
\item {\bf Experiment statistical significance}
    \item[] Question: Does the paper report error bars suitably and correctly defined or other appropriate information about the statistical significance of the experiments?
    \item[] Answer: \answerYes{}
    \item[] Justification: The main result tables report mean $\pm$ standard deviation for learned rows, Section~\ref{sec:experiments} defines what those summaries capture in each regime, and Appendix Table~\ref{tab:transfer_significance} reports paired Wilcoxon signed-rank tests across the 13 transfer folds for the comparisons that support the main transfer claims.
\item {\bf Experiments compute resources}
    \item[] Question: For each experiment, does the paper provide sufficient information on the computer resources (type of compute workers, memory, time of execution) needed to reproduce the experiments?
    \item[] Answer: \answerYes{}
    \item[] Justification: Appendix~\ref{app:training} reports the hardware used, data-loader configuration, aggregate runtime for the reported experiments, and the observed runtime range across model families.
\item {\bf Code of ethics}
    \item[] Question: Does the research conducted in the paper conform, in every respect, with the NeurIPS Code of Ethics \url{https://neurips.cc/public/EthicsGuidelines}?
    \item[] Answer: \answerYes{}
    \item[] Justification: The paper includes an \emph{Ethics, Licensing, and Release} section, states the animal-protocol approval, and describes the release scope and licensing of the dataset and code.
\item {\bf Broader impacts}
    \item[] Question: Does the paper discuss both potential positive societal impacts and negative societal impacts of the work performed?
    \item[] Answer: \answerYes{}
    \item[] Justification: The \emph{Ethics, Licensing, and Release} section states the intended research use of the benchmark, explains that no material societal risks are currently foreseen, and notes only a remote surveillance-style misuse scenario for related attribution or localization systems.
\item {\bf Safeguards}
    \item[] Question: Does the paper describe safeguards that have been put in place for responsible release of data or models that have a high risk for misuse (e.g., pre-trained language models, image generators, or scraped datasets)?
    \item[] Answer: \answerYes{}
    \item[] Justification: The release is an animal-behavior benchmark with no personally identifiable information: birds are referenced only by experiment-local backpack color codes, no human subjects are recorded, and no owner or facility metadata is shipped. All data were collected under Cantonal Veterinary Office of Zurich license ZH148/2020 (see \emph{Ethics, Licensing, and Release}), the dataset is distributed under CC-BY-4.0, and the release scope (curated verified single-vocalizer events plus the V33DA++ extension) excludes the raw continuous recordings that would be needed for surveillance-style re-use.
\item {\bf Licenses for existing assets}
    \item[] Question: Are the creators or original owners of assets (e.g., code, data, models), used in the paper, properly credited and are the license and terms of use explicitly mentioned and properly respected?
    \item[] Answer: \answerYes{}
    \item[] Justification: The paper credits the external software and pretrained components it uses and states in the \emph{Ethics, Licensing, and Release} section that these assets remain subject to their original licenses and terms of use.
\item {\bf New assets}
    \item[] Question: Are new assets introduced in the paper well documented and is the documentation provided alongside the assets?
    \item[] Answer: \answerYes{}
    \item[] Justification: The paper releases a new dataset and codebase, describes their scope and limitations in the main text, and points to the public release locations and appendix documentation.
\item {\bf Crowdsourcing and research with human subjects}
    \item[] Question: For crowdsourcing experiments and research with human subjects, does the paper include the full text of instructions given to participants and screenshots, if applicable, as well as details about compensation (if any)? 
    \item[] Answer: \answerNA{}
    \item[] Justification: The work does not involve crowdsourcing or research with human subjects. The manual cross-modal validation of candidate events described in Section~\ref{sec:dataset} was performed by the first author as part of the research contribution, not by external annotators, so no participant instructions or compensation apply.
\item {\bf Institutional review board (IRB) approvals or equivalent for research with human subjects}
    \item[] Question: Does the paper describe potential risks incurred by study participants, whether such risks were disclosed to the subjects, and whether Institutional Review Board (IRB) approvals (or an equivalent approval/review based on the requirements of your country or institution) were obtained?
    \item[] Answer: \answerNA{}
    \item[] Justification: The research does not involve human subjects.
\item {\bf Declaration of LLM usage}
    \item[] Question: Does the paper describe the usage of LLMs if it is an important, original, or non-standard component of the core methods in this research? Note that if the LLM is used only for writing, editing, or formatting purposes and does \emph{not} impact the core methodology, scientific rigor, or originality of the research, declaration is not required.
    \item[] Answer: \answerNA{}
    \item[] Justification: LLMs are not a component of the benchmark construction, models, or experimental methodology reported in the paper; any LLM assistance in prose editing did not contribute to dataset design, model design, experimental protocol, or result interpretation.
\end{enumerate}

\end{document}